\documentclass[12pt,letterpaper]{article}
\usepackage{hyperref}
\usepackage{latexsym}
\usepackage{amssymb,amsfonts,amsmath}
\usepackage{graphicx}
\usepackage{indentfirst}
\usepackage{amsmath}
\usepackage{amsfonts}
\usepackage{amssymb}%
\usepackage{comment}
\usepackage{mathtools}
\usepackage{amsthm}
\usepackage{enumerate}
\ifx\pdfoutput\undefined
\else
\fi
\hypersetup{colorlinks=false,bookmarksopen,bookmarksnumbered,citecolor=blue,
pdfstartview=FitH}

\parskip=\medskipamount

\def\a{\alpha}

\def\b{\beta}

\def\d{\delta}
\def\e{\epsilon}

\def\g{\gamma}
\def\G{\Gamma}

\def\k{\kappa}

\def\s{\sigma}

\def\S{\Sigma}

\def\tr{{\rm tr}}

\newcommand{\pa}{\partial}

\newcommand{\sect}[1]{\section{#1}\setcounter{equation}{0}}

\newcommand{\be}{\begin{equation}}
\newcommand{\ee}{\end{equation}}
\newcommand{\bea}{\begin{eqnarray}}
\newcommand{\eea}{\end{eqnarray}}

\newcommand{\ba}{\begin{array}}
\newcommand{\ea}{\end{array}}

\def\double #1{#1{\hbox{\kern-2pt $#1$}}}

\newcommand{\bsubeq}{\begin{subequations}}
\newcommand{\esubeq}{\end{subequations}}

\newcommand{\virgolette}{``}
 
\theoremstyle{definition}

\theoremstyle{plain}

\begin{document}

\begin{titlepage}
\begin{flushright}
\par\end{flushright}
\vskip 0.5cm
\begin{center}
\textbf{\huge \bf  Surface Operators in Superspace}
\vskip 1cm
\vskip 0.5cm
\large {\bf C.~A.~Cremonini}$^{~a,b,}$\footnote{carlo.alberto.cremonini@gmail.com}, 
\large {\bf P.~A.~Grassi}$^{~c,d,e,}$\footnote{pietro.grassi@uniupo.it}, 
and
\large {\bf S.~Penati}$^{~f,}$\footnote{silvia.penati@mib.infn.it}, 
\vskip .5cm {
\small
\medskip
\centerline{$^{(a)}$ \it Dipartimento di Scienze e Alta Tecnologia (DiSAT),}
\centerline{\it Universit\`a degli Studi dell'Insubria, via Valleggio 11, 22100 Como, Italy}
\medskip
\centerline{$^{(b)}$ \it INFN, Sezione di Milano, via G.~Celoria 16, 20133 Milano, Italy} 
\medskip
\centerline{$^{(c)}$
\it Dipartimento di Scienze e Innovazione Tecnologica (DiSIT),} 
\centerline{\it Universit\`a del Piemonte Orientale, viale T.~Michel, 11, 15121 Alessandria, Italy}
\medskip
\centerline{$^{(d)}$
\it INFN, Sezione di Torino, via P.~Giuria 1, 10125 Torino, Italy}
\medskip
\centerline{$^{(e)}$
\it Arnold-Regge Center, via P.~Giuria 1,  10125 Torino, Italy}
\medskip
\centerline{$^{(f)}$ 
\it Dipartimento di Fisica, Universit\`a degli studi di Milano-Bicocca, }
\medskip
\centerline{\it and INFN, Sezione di Milano-Bicocca, Piazza della Scienza 3, 20126 Milano, Italy}
\medskip
}
\end{center}
\vskip  0.1cm
\begin{abstract}
We generalize the geometrical formulation of Wilson loops recently introduced in \cite{Cremonini:2020mrk} to the description of Wilson Surfaces. For N=(2,0) theory in six dimensions, we provide an explicit derivation of BPS Wilson Surfaces with non-trivial coupling to scalars, together with their manifestly supersymmetric version. We derive explicit conditions which allow to classify these operators in terms of the number of preserved supercharges. We also discuss kappa-symmetry and prove that BPS conditions in six dimensions arise from kappa-symmetry invariance in eleven dimensions. Finally, we discuss super-Wilson Surfaces - and higher dimensional operators - as objects charged under global $p$-form (super)symmetries generated by tensorial supercurrents. To this end, the construction of conserved supercurrents in supermanifolds and of the corresponding conserved charges is developed in details. 
\end{abstract}
\vfill{}
\vspace{1.5cm}
\end{titlepage}
\newpage\setcounter{footnote}{0}
\tableofcontents

\section{Introduction} \setcounter{equation}{0}

In gauge theories an important physical quantity is the Wilson loop defined as the holonomy of the gauge connection along a one-dimensional contour. This has a natural generalization to higher dimensional contours, whenever the theory lives in $D>3$ dimensions and contains higher-rank tensor fields.  In particular, Wilson Surfaces (WS) are defined in terms of a surface integral of a two-form tensor \cite{Ganor:1996nf}
\be\label{intro:1}
W[\s] = e^{\G[\s]}, \quad {\rm with} \quad \G[\s] = \int_\s B_{\mu\nu} dx^\mu dx^\nu
\ee
where $\s$ is a two-dimensional submanifold. Notable examples are WS in four dimensions with $B_{\mu\nu}$ identified with the abelian gauge field strength (electric surface operator), WS in the six-dimensional $N=(2,0)$ theory living on one M5-brane where $B_{\mu\nu}$ is the self-dual two-form of the tensor multiplet, and WS in ten or eleven dimensions.
In topological theories their expectation values are expected to yield invariants of higher-dimensional knots \cite{Cattaneo:2002tk}. 

In $N=4$ SYM theory, general classes of surface operators which support some supersymmetry (BPS WS) have been first classified in \cite{Gukov:2006jk,Gukov:2008sn,Gukov:2014gja}. They naturally arise in string theory as configurations of intersecting D3-branes \cite{Constable:2002xt,Gukov:2006jk,Koh:2009cj,Gukov:2014gja} or fractional D3-branes in orbifold backgrounds \cite{Ashok:2020ekv,Ashok:2020jgb}. A more general class of supersymmetric surface operators in $N=4$ SYM can be obtained as holographic duals of D3/D7-brane intersections \cite{Buchbinder:2007ar}. 
 
Similarly, in the context of the AdS$_7$/CFT$_6$ correspondence, BPS WS of the $N=(2,0)$ superconformal field theory living on one M5-brane have a dual description in terms of intersecting M5/M2-branes \cite{Maldacena:1998im}. Explicit M5-brane string soliton solutions in AdS$_7 \times {\rm S}^4$ background have been found, which correspond to the expectation value of BPS surface operators, in different representations  and different surface topologies \cite{Lunin:2007ab,Chen:2007ir,Chen:2007tt,Agarwal:2018tso}. Their field theory realization is given in terms of a generalized 2-form which is a linear combination of the 2-form and the five scalar fields of the 6D tensor multiplet \cite{Gustavsson:2004gj,Mezei:2018url,Drukker:2020dcz}. This construction can be understood either by generalizing the formulation of BPS Wilson loops or, alternatively, in a dimensional reduction approach from M-theory. 

Correlation functions of WS's, and WS with local operators, Wilson and 't Hooft loops in four \cite{Drukker:2008wr,Tan:2013yza,Bianchi:2019sxz} and six \cite{Corrado:1999pi,Drukker:2020swu} dimensions have been computed, perturbatively or using (super)conformal techniques or the dual supergravity description. The dual description has also been used to compute the Operator Product Expansion of surface operators in the large $N$ limit \cite{Berenstein:1998ij,Corrado:1999pi,Chen:2007zzr}.

Surface operators exhibit a conformal anomaly \cite{Graham:1999pm}, as expected for all even dimensional defects. The anomaly in six dimensions has been determined perturbatively \cite{Henningson:1999xi,Gustavsson:2003hn,Gustavsson:2004gj,Drukker:2020dcz} and holographically \cite{Young:2011aa,Drukker:2020swu}, and the results are consistent with what has been obtained from the entanglement entropy for the bubbling M5/M2 geometry \cite{Estes:2018tnu,Jensen:2018rxu}, and exactly from the determination of the corresponding superconformal index \cite{Chalabi:2020iie}.

Superconformal surface defects in four dimensional $N \geq 2$ superconformal theories have been used to probe the low energy phases of the four dimensional theory.
The low-energy dynamics of the two-dimensional theory living on a superconformal defect is featured by a twisted superpotential which controls its interactions with the bulk degrees of freedom. This 2d/4d system has been extensively investigated by exploiting superconformal techniques, the associated chiral algebra modules \cite{Gaiotto:2009fs,Gaiotto:2013sma,Cordova:2017ohl,Cordova:2017mhb} and equivariant localization \cite{Ashok:2017lko,Ashok:2018zxp}. A bootstrap program for solving conformal field theories with conformal defects has been initiated in \cite{Liendo:2012hy} and extensively developed along the years (a non-exhaustive list of references includes \cite{Gaiotto:2013nva,Gliozzi:2015qsa,Gliozzi:2016cmg,Liendo:2016ymz,Bissi:2018mcq,Kaviraj:2018tfd}).

From a QFT perspective, interest in surface operators is triggered by the fact that they describe dynamical defects that are charged under two-form global symmetries \cite{Gaiotto:2014kfa,Cordova:2018cvg,Seiberg:2019vrp}. Moreover, in a continuum field theory description of fractons and lineons \cite{Pretko:2016kxt, Pretko:2018jbi, Gromov:2018nbv, Seiberg:2020bhn,Seiberg:2020wsg,Seiberg:2020cxy}, they arise as the gauge-invariant phases developed by these observables in their constrained motion. 

Within the general context of supersymmetric theories, in this paper we introduce superWS - that is the manifestly supersymmetric version of operators of the form \eqref{intro:1} - and study their properties and invariances. This is carried out using supergeometry, basically rephrasing what has been done in \cite{Cremonini:2020mrk} for super-Wilson loops. We first reformulate expression \eqref{intro:1} as an integral on the entire manifold by making use of a Poincar\'e dual which localizes the integral on the surface. Then we extend this definition to a generic supermanifold endowed with super-Poincar\'e duals that localize integrals to supersurfaces and write the supersymmetric version of $\G$ in \eqref{intro:1} as the integral of a top form on the entire supermanifold (see eq. \eqref{AKAF}). This has the advantage to make the superWS manifestly invariant under superdiffeomorphisms, from which constraints for supersymmetry and kappa-symmetry invariances easily follow, as we discuss in details. 

In particular, we focus on superWS in the six dimensional $N=(2,0)$ superconformal theory, whose tensor supermultiplet contains a self-dual superform $B^{(2|0)}$ suitable for defining superWS. We study both ordinary superWS defined only in terms of $B^{(2|0)}$ and generalized superWS which contain also couplings to scalar multiplets. The two kinds of operators simply differ by a $d$-exact shift of the corresponding  super-Poincar\'e duals. For both kinds of operators we determine 
the constraining equations which select the class of supersurfaces leading to BPS superWS. This class includes the 1/2 BPS solutions corresponding to planar and spherical surfaces already discussed in the literature. More generally, we find that in Minkowski spacetime there are no 1/2 BPS WS corresponding to spacelike surfaces. This result resembles what is already known for Wilson loops \cite{Ouyang:2015ada}.  Similarly, we determine the conditions which ensure kappa-symmetry invariance of superWS. Remarkably, they coincide with the conditions for a static supermembrane to be kappa-symmetric \cite{Berkovits:2002uc}. 

As WS are objects charged under two-form tensor symmetries, we expect superWS to be objects charged under symmetries generated by {\em tensorial supercurrents}. Aimed at investigating this question, in the last part of the paper we construct conserved tensorial supercurrents and the corresponding conserved supercharges within the framework of supergeometry.  These are the supersymmetric generalization of the tensorial currents introduced in \cite{Gaiotto:2014kfa}. We then prove that WS and higher dimensional operators are indeed the charge carriers for these symmetries.  

The plan of the paper is the following. For reader's convenience, in section 2 we briefly summarise the main concepts of supergeometry that are used along the paper.  As a warming-up, in section 3 we rephrase ordinary Wilson Surfaces in terms of integrals on the entire manifold of 2-form fields times the PCO which localizes the integral on a given surface. Section 4 contains our main proposal, that is the general expression for a superWS in superspace (see eq. \eqref{BKC}). We stick to the abelian case, as non-abelian WS are not well understood yet, though various proposals for generalizing the concept of normal-ordered exponential have already appeared in the literature \cite{Chepelev:2001mg,Hofman:2002ey,Baez:2010ya,Ho:2012nt,Kim:2019owc}. Its behaviour under supersymmetry and kappa-symmetry depends entirely on the behaviour of the supersurface, as we discuss in details in sections 5 and 6 for the six dimensional $N=(2,0)$ theory. Section 7 is devoted to the study of tensorial conservation laws in supermanifolds, and to the interpretation of superWS and their higher dimensional super-cousins as the objects charged under the global symmetries generated by these supercurrents. Finally, section 8 is devoted to some conclusions and perspectives. In particular, we address the fact that our construction opens the possibility of studying a continuum theory for fractons and lineons in superspace ({\em superfractons} and {\em superlineons}), as we will discuss in a forthcoming paper \cite{new}. Four appendices follow, one summarising our conventions in six dimensions, one recalling basic definitions about the Hodge operator in supermanifolds, one including an alternative discussion of conservation laws that makes use of an explicit surface parametrization, and finally one where we define a supersymmetric version of the linking number between two supersurfaces, required to define the action of charge operators on superWS and higher dimensional objects.

\section{Supergeometry and Picture Changing Operators} \setcounter{equation}{0}

In this section we briefly review the basics of geometry on supermanifolds and the role of Picture Changing Operators. A more extensive discussion on supergeometry can be found in \cite{Belo,Witten:2012bg, CatenacciGrassiNoja}, whereas applications of this formalism to field theories have been developed in \cite{Castellani:2014goa, Catenacci:2016qzd, CatenacciGrassiNoja2, CremoniniGrassi, CremoniniGrassi2}. A geometric formulation of (super) Wilson Loops has been recently proposed in \cite{Cremonini:2020mrk}.

A supermanifold $\mathcal{SM}$ of dimensions $(n|m)$ is locally described by a set of $n$ even variables $\lbrace x^a \rbrace_{a=1, \dots, n}$ and a set of $m$ odd variables $\lbrace \theta^\alpha \rbrace_{\alpha=1, \dots, m}$. On supermanifolds it is possible to develop the full Cartan calculus. The basic ingredients are super-differential forms (henceforth \emph{superforms}), that is elements of the cotangent bundle $T^* \mathcal{SM}$ expanded on a basis of odd $\lbrace dx^a \rbrace$ and even $\lbrace d \theta^\alpha \rbrace$ differential forms. Since the $d \theta$'s are commuting quantities, there is no notion of \emph{top form} in the complex of superforms. The notion of top form has to be found into a new complex of forms, known as \emph{integral forms}. We define such objects by following the strategy of Belopolsky \cite{Belo}, where integral forms are distributional-like forms. In particular, forms on supermanifolds are described by the usual form number $p$ and by a new grading $q$ called \emph{picture number}.

Precisely, a generic superform in $\mathcal{SM}$ locally reads
\begin{eqnarray} \label{PNAA}
\hspace{-0.6cm} \omega^{(p|0)} (x, \theta, dx, d\theta)  = \omega(x,\theta) \, dx^{a_1}\dots dx^{a_r} \, \left( d\theta^{\a_1} \right)^{g_1} \dots \left( d\theta^{\a_s} \right)^{g_s}
\end{eqnarray}
where $p= r + \sum_{i=i}^s g_i$ and $q=0$. By contrast, a generic \emph{integral form} is written as
\begin{equation} \label{PNAD}
	\omega^{(p|m)} (x, \theta, dx, d\theta)  = \omega(x,\theta) \, dx^{a_1}\dots dx^{a_r} \, \delta^{(b_1)} (d\theta^{\beta_{1}}) \dots  \delta^{(b_{m})}(d\theta^{\beta_{m}})
\end{equation}
with $p= r   \!-\! \sum_{j=i}^m b_j$ and $q=m$. In this expression there are no $d \theta$'s, due to the following distributional properties
\begin{equation} \label{PNAE}
	d \theta^\alpha \delta \left( d \theta^\alpha \right) = 0 \ , \quad d \theta^\alpha \delta^{(p)} \left( d \theta^\alpha \right) = - p \delta^{(p-1)} \left( d \theta^\alpha \right) 
\end{equation}
In order to keep track of the orientation of the supermanifold we require $\delta \left( d \theta \right)$ to be odd objects, hence $m$ is the maximum number of $\delta$'s that may appear in a given form. We note that the \emph{picture number} counts indeed the number of $\delta$'s appearing in a given form. While superforms have $q=0$, integral forms have maximal picture number $q=m$. 

The notion of \emph{top form} used for integration in supergeometry is contained in the space $\Omega^{(n|m)} \left( \mathcal{SM} \right)$, called \emph{Berezinian bundle}, analogous to the determinant bundle of usual geometry. A generic top integral form $\omega^{(n|m)} \in \Omega^{(n|m)} \left( \mathcal{SM} \right)$ reads
\begin{equation}
	\omega^{(n|m)} = f_{[i_1 \ldots i_n ][\alpha_1 \ldots \alpha_m]} (x, \theta) dx^{i_1} \ldots dx^{i_n} \delta \left( d \theta^{\alpha_1} \right) \ldots \delta \left( d \theta^{\alpha_m} \right) = f(x,\theta) d^nx \delta^m \left( d \theta \right)
\end{equation}
As in  conventional geometry, we can define the integral of a top form on a superspace endowed with a super-measure $[dx d\theta d(dx) d(d\theta)]$, invariant under coordinate transformations. Following \cite{Witten:2012bg}, we write
 \begin{eqnarray}
\label{PNAF}
I[\omega] = \int_{{\cal SM}} \omega^{(n|m)} = 
\int_{T^* {\cal SM}} \omega^{(n|m)}(x, \theta, dx, d\theta) \;  [dx d\theta d(dx) d(d\theta)]
\end{eqnarray}
where $dx$ and $d(d\theta)$ define Lebesgue integrals, while the integrations over $d\theta$ and $d(dx)$ are Berezin integrals. 

\vskip 10pt
Picture Changing Operators (PCOs) are conveniently introduced to define the integration of generic superforms in a supermanifold. Let us consider for example a bosonic submanifold $\mathcal{N} \subset \mathcal{SM}$, with $\text{dim} \, \mathcal{N} = (p|0)$, defined by the embedding $\iota : \mathcal{N} \to \mathcal{SM}$, and a $(p,0)$ superform $\omega^{(p|0)}$. We define the integration of $\omega^{(p|0)}$ on $\mathcal{N}$ as
\begin{equation} \label{PCOAA}
	\int_{\cal N} \iota_*\omega^{(p|0)} \equiv \int_{\cal SM} \omega^{(p|0)} \wedge \mathbb{Y}^{(n-p|m)}_{\cal N}
\end{equation}
where $\iota_*$ is the pull-back map and $\mathbb{Y}^{(n-p|m)}_{\cal N}$ is the Poincar\'e dual of the embedding $\iota$, called \emph{Picture Changing Operator}, from string theory literature (see e.g. \cite{Deligne}). The PCO in \eqref{PCOAA} is independent of the surface parametrization, it only depends on the embedding through its homology class and satisfies the closure, but non-exactness conditions
\begin{equation}
	d \mathbb{Y}^{(n-p|m)}_{\cal N} = 0 \ , \qquad \mathbb{Y}^{(n-p|m)}_{\cal N} \neq d \Sigma^{(n-p-1|m)}
\end{equation}

When the submanifold ${\cal N}$ is one-dimensional and $\omega^{(1|0)}$ is a gauge connection, equation \eqref{PCOAA} provides a geometric construction of (super) Wilson loops \cite{Cremonini:2020mrk}. Many properties of the Wilson operator, like supersymmetry and kappa invariance, are dictated by the behavior of the PCO. 

Changing the submanifold $\mathcal{N}$ to an homologically equivalent one $\mathcal{N}'$, amounts to changing the PCO by the addition of a $d$-exact term
\begin{equation}\label{PCOAB}
\mathbb{Y}^{(n-p|m)}_{\cal N} \to \mathbb{Y}^{(n-p|m)}_{\cal N'} = \mathbb{Y}^{(n-p|m)}_{\cal N} + d \Omega^{(n-p-1|m)}
\end{equation}
This mechanism has been extensively used in the context of (super) Wilson loops, in particular for tuning the amount of supersymmetry preserved by a given operator \cite{Cremonini:2020mrk}.  As a remarkable example, it has been proved that the generalized Wilson-Maldacena holonomy \cite{Maldacena:1998im}  which leads to a BPS operator in $N=4$ SYM theory can be generated from an ordinary non-BPS holonomy by a $d$-exact shift of the corresponding PCO. In the present paper, we are going to generalize this  mechanism to the case of Wilson Surfaces.

\section{Wilson Surfaces} \setcounter{equation}{0}
\label{sect:bosonicWS}

We begin by considering Wilson Surfaces (WS) in ordinary manifolds with no supersymmetry. Given an $n$-dimensional manifold ${\mathcal M}$ and a two-dimensional surface $\sigma$ immersed in it, a WS corresponding to a $2$-form $B^{(2)}$ on ${\mathcal M}$ is defined as 
\begin{eqnarray}
\label{WSA}
W_S[\sigma] = e^{ \Gamma } \,, ~~~~~ \Gamma =  \int_{\sigma} \iota_* B^{(2)} 
\end{eqnarray}
where $\iota$ is the immersion of $\sigma$ into ${\mathcal M}$ (see for instance \cite{Gukov:2014gja} for a review on surface operators in four-dimensional gauge theories). 

Generalizing the geometric construction of Wilson loops introduced in \cite{Cremonini:2020mrk}, we reformulate this definition in terms of a suitable PCO which encodes all the geometric data featuring $W_S$. If the surface is defined by the set of equations $\phi_i(x^a)=0$, with $i= 1, \dots , n-2$, we introduce the PCO 
\begin{eqnarray}
{\mathbb Y}^{(n-2)}_\s = \prod_{i=1}^{n-2} \d(\phi_i) d\phi_i
\end{eqnarray}
dual to the immersion of the two-dimensional surface into the manifold. Therefore, the WS exponent in \eqref{WSA} can be rewritten as a top form integrated on the whole manifold \cite{Gukov:2006jk}
\begin{eqnarray}\label{WSAbis}
\Gamma =  \int_{\cal M}  B^{(2)} \wedge {\mathbb Y}^{(n-2)}_\s
\end{eqnarray}
Thanks to the $d$-closure of the PCO this expression is manifestly invariant under $U(1)$ gauge transformations $B \to B + d\beta$, as long as $\s$ is a compact surface with no boundaries, or fields vanish at the intersection $\s \cap \partial {\cal M}$. Moreover, under a smooth deformation of the surface the PCO changes by an exact term, ${\mathbb Y}^{(n-2)}_\s \to {\mathbb Y}^{(n-2)}_\s + d \eta^{n-3}$, and correspondingly $\G$ varies as
\begin{eqnarray}
\G \to \G + \int_{\cal M}  H^{(3)} \wedge \eta^{n-3}
\end{eqnarray}
where $H^{(3)} = d B^{(2)}$ is the field strength.

Expression \eqref{WSAbis} can be recast in a more familiar form. To this end, we consider the parametrisation of the surface $\sigma$ in \eqref{WSA} by two real parameters $(z, \bar z) \rightarrow \{ x^a(z, \bar z)\}$, with $(z, \bar z) \in {\Delta}  \subseteq {\mathbb R}^2$. 
We then define the enlarged manifold ${\cal M} \times \Delta$ described by  coordinates $(x^a, z, \bar z)$ and construct the PCO dual to the embedding $(z, \bar z) \rightarrow (x^a(z, \bar z), z, \bar z)$ as follows
\begin{eqnarray}
\label{AKACA}
{\mathbb Y}^{(n)}_\sigma &=& \prod_{a=1}^n \delta\Big(x^a - x^a(z, \bar z) \Big) \bigwedge_{a=1}^n (dx^a - \partial_z x^a dz - \partial_{\bar z} x^a d\bar z )  \nonumber \\
 &=&\prod_{a=1}^n \delta\Big(x^a - x^a(z, \bar z) \Big)  \left[
 \left( \bigwedge_{a=1}^n dx^a  + \sum_{b=1}^n (-1)^b  (\partial_z x^b dz 
+ \partial_{\bar z} x^b d\bar z) \bigwedge_{a\neq b} dx^a \right) \right.
 \nonumber \\
 &~& \quad \left. + \sum_{b=1,c=1}^n (-1)^{b+c}  \left(\partial_z x^b   \partial_{\bar z} x^c dz \wedge d\bar z  \bigwedge_{a\neq b, a\neq c} dx^a
 \right) \right]
\end{eqnarray} 
Multiplying by the $2$-form parametrized as $B^{(2)} = B_{ab} \, dx^a \wedge dx^b$, it is easy to see that only the last term survives and we end up with
\begin{eqnarray}
\label{AKACB}
B^{(2)} \wedge {\mathbb Y}^{(n)}_\sigma
 = B_{bc} \partial_z x^b \partial_{\bar z}x^c  \, dz\wedge d\bar z \, \prod_{a=1}^n  \delta\Big(x^a - x^a(z, \bar z) \Big)  \bigwedge_{a=1}^n dx^a 
\end{eqnarray}
Integrating on ${\cal M}\times \Delta$ we finally obtain
\begin{eqnarray}
\label{AKACC}
\int_{{\cal M}\times \Delta} \!\!\!\! \!\!\!\! B^{(2)} \wedge {\mathbb Y}^{(n)}_\s  = 
\int_\sigma    B_{ab}(x(z, \bar z)) \, \partial_z x^a \partial_{\bar z}x^b \, dz d\bar z
\end{eqnarray}
which coincides with $\G$ in \eqref{WSA} when $\sigma$ is parametrized by coordinates $(z, \bar z)$. Therefore, an alternative way to define a WS is 
\begin{equation}\label{WSB}
W_S[\sigma] = e^{ \Gamma } \,, ~~~~~  \G = \int_{{\cal M}\times \Delta} \!\!\!\!  \!\!\!\!  B^{(2)} \wedge {\mathbb Y}^{(n)}_\s
\end{equation}

The geometric formulation of WS has many advantages. The main one is that the integral is extended to the whole manifold, the $2$-form $B^{(2)}$ is generically defined on it rather being constrained to live on the surface,  while all the information regarding the surface is totally encoded in the PCO. This makes the study of $W_S[\sigma]$ invariances much easier. In particular, expression \eqref{WSB} is manifestly invariant under diffeomorphisms of the manifold. The implications of this property will be further investigated in the rest of the paper.

\section{Super Wilson Surfaces} \setcounter{equation}{0}\label{sect:susyWS}

Among the many advantages of formulation \eqref{WSAbis} (or \eqref{WSB}) for WS we count the fact that it allows for a straightforward generalisation to supermanifolds, thus leading to an easy definition of super Wilson surfaces. 

We consider a supermanifold ${\cal SM}$ described by coordinates $Z^M \equiv (x^a, \theta^\a)$, with $a=1,\dots ,n$, $\a = 1, \dots , m$, and assign a super $2$-form $B^{(2|0)}$ on it. If $\S$ is a dimension-$(2|0)$ supersurface whose immersion in ${\cal SM}$ is described by the PCO ${\mathbb Y}^{(n-2|m)}_\Sigma$,  
 we define the superWS as
\begin{eqnarray}
\label{AKAF}
W_S[\Sigma] = e^{ \Gamma } \,, ~~~~~  \Gamma = \int_{\cal SM} \!\!\!\!  B^{(2|0)} \wedge {\mathbb Y}^{(n-2|m)}_\Sigma
\end{eqnarray}
The $2$-form is generically defined in ${\cal SM}$, while the geometrical data featuring the supersurface are entirely captured by the PCO. This is factorized into a bosonic and a fermionic part, ${\mathbb Y}^{(n-2|m)}_\Sigma = {\mathbb Y}^{(n-2|0)}_\Sigma \times {\mathbb Y}^{(0|m)}_\Sigma$, where ${\mathbb Y}^{(n-2|0)}_\Sigma$ localizes the bosonic coordinates on $\Sigma$ whereas ${\mathbb Y}^{(0|m)}_\Sigma$ localizes the fermionic ones. If the supersurface is defined by means of algebraic equations, the PCO  ${\mathbb Y}^{(n-2|m)}_\Sigma$ is the product of the Dirac delta functions localizing on that surface.

This  expression for $\Gamma$ can be made more explicit if we parametrize the supersurface $\Sigma$ in terms of smooth functions $ (z, \bar z) \to Z^M(z, \bar z)$ on 
$\Delta \subseteq {\mathbb R}^2$. For the bosonic part of the PCO we can proceed exactly as done in section \ref{sect:bosonicWS}, by including $(z, \bar z)$ 
as extra bosonic coordinates and extending the integration to the supermanifold ${\cal SM} \times \Delta$. A straightforward supersymmetrization of eq. \eqref{AKACA} leads to 
\begin{eqnarray}\label{BKA2}
{\mathbb Y}^{(n|0)}_\Sigma = \prod_{a=1}^n \delta\Big(x^a - x^a(z, \bar z) \Big) \bigwedge_{a=1}^n (V^a - \Pi^a_z(z, \bar z) dz - \Pi^a_{\bar z}(z, \bar z) d\bar z) 
\end{eqnarray}
where we have defined $V^a = dx^a + \theta \g^a d\theta$, $\Pi^a_z(z, \bar z)  = (\partial_z{x}^a +  \theta \gamma^a \partial_z{\theta})$ and 
$\Pi^a_{\bar z}(z, \bar z)  = (\partial_{\bar z}{x}^a +  \theta \gamma^a \partial_{\bar z}{\theta})$. 

For the PCO of the fermionic sector we choose
\begin{eqnarray}
\label{BKA}
{\mathbb Y}^{(0|m)}_\S &=& \prod_{\alpha=1}^m \Big(\theta^\a - \theta^\a(z, \bar z)\Big) \delta\Big(\psi^\a -(\partial_z \theta^\b(z, \bar z) dz + \partial_{\bar z} \theta^\b(z, \bar z) d\bar z) \Big) \\
&=& \prod_{\alpha=1}^m \Big(\theta^\a - \theta^\a(z, \bar z)\Big)  \left[
\Big(1 - \sum_\beta (\partial_z \theta^\b(z, \bar z) dz + \partial_{\bar z} \theta^\b(z, \bar z) d\bar z) \iota_\b  \right. \nonumber \\
&~& \quad \left. + 
\sum_{\beta,\gamma} (\partial_z \theta^\b(z, \bar z) \partial_{\bar z} \theta^\gamma(z, \bar z) dz d\bar z) \iota_\b \iota_\gamma
\Big) \prod_{\alpha=1}^m \delta(\psi^\a) \right] \nonumber
\end{eqnarray}
where $\psi^\a = d\theta^\a$ and 
in the second line we have expanded the Dirac delta functions exploiting the presence of the anticommuting one-forms $dz$ and $d\bar z$. 
Here $ \iota_\b$ is the contraction along the supercovariant derivative $D_\beta$. Using a shorter notation we can then write
\begin{eqnarray}\label{BKA3}
{\mathbb Y}^{(n|m)}_\S& \equiv&  {\mathbb Y}^{(n|0)}_\S \wedge {\mathbb Y}^{(0|m)}_\S  = \\
&=&   \d^{(n)} (x - x(z, \bar z)) \, ( V - \Pi_z dz - \Pi_{\bar z} d\bar z)^n \, \wedge \, (\theta - \theta(z, \bar z) )^m \, \d^{(m)}(\psi - \partial_z{\theta} dz - \partial_{\bar z} \theta d\bar z ) \nonumber
\end{eqnarray}
The exponent $\G$ in eq. \eqref{AKAF} can then be rewritten as
\begin{equation}\label{gamma}
\G = \int_{{\cal SM} \times \Delta} \!\!\!\! \!\!\!\!  B^{(2|0)} \wedge {\mathbb Y}^{(n|m)}_\Sigma
\end{equation}

We now elaborate on this expression. Expanding $B^{(2|0)}$ in terms of the superspace geometric objects, and focusing first on the fermionic part of the PCO, we can write
\begin{eqnarray}
\label{BKB}
 \hspace{-0.8cm} \Gamma &=& 
 \int_{{\cal SM} \times \Delta} \Big(B_{ab}(x, \theta) V^a V^b + 
B_{a\beta}(x, \theta) V^a \psi^\b + B_{\a\beta}(x, \theta) \psi^\a \psi^\b  
\Big) 
\wedge 
\prod_{\alpha=1}^m \Big(\theta ^\a- \theta^\a(z, \bar z)\Big) \times \nonumber \\
&&\Big(1 - \sum_\beta (\partial_z\theta^\b dz + \partial_{\bar z}\theta^\b d\bar z)\iota_\b + \sum_{\beta,\gamma} 
(\partial_z\theta^\b \partial_{\bar z}\theta^\gamma dz d\bar z )
\iota_\b \iota_\gamma\Big) \prod_{\alpha=1}^m \delta(\psi^\a) \; \wedge {\mathbb Y}^{(n|0)}_\Lambda 
\end{eqnarray}
where $B_{ab}$ and $B_{\a\b}$ are bosonic components, whereas $B_{a\a}$ are fermionic.
Now, due to the presence of the factor $\prod_\a \delta(\psi^\a)$ the only non-vanishing contributions come from terms in the integrand which do not contain any power of $\psi^\a$, like for instance $B_{ab}(x, \theta) dx^a dx^b$ from the first term, or terms linear and quadratic in $\psi^\a$ on which the action of the contraction(s) $\iota_\a$ has the effect of replacing $\psi^\a \to \partial_z \theta^\a dz + \partial_{\bar z} \theta^\a d\bar z$. Therefore, using PCO \eqref{BKA2} to localize also the bosonic coordinates on the supersurface $\S$, from the previous equation we easily find
\begin{eqnarray}
\label{BKC}
\Gamma \! 
&=& \int_\Sigma   \Big(B_{ab} \Pi^a_z \Pi^b_{\bar z} \! + \! 
B_{a\a}  (\Pi^a_z \partial_{\bar z}\theta^\a + \Pi^a_{\bar z} \partial_{z}\theta^\a) + 
B_{\a\b} \partial_{z}\theta^\a \partial_{\bar z}\theta^\beta\Big) dz d\bar z
\end{eqnarray}
This equation provides the supersymmetric version of the WS in \eqref{AKACC}. In fact, if we choose $\S$ to be an ordinary two-dimensional surface localised at $\theta_\alpha(z ,\bar{z}) = 0$, the corresponding PCO reduces to ${\mathbb Y}^{(n|0)}_\S \wedge {\mathbb Y}^{(0|m)}_0$ with
\begin{equation}\label{BKD}
{\mathbb Y}^{(0|m)}_0 = \theta^m \d^{(m)} (\psi)
\end{equation}
and the integral in \eqref{BKC} coincides with \eqref{AKACC}. 

Although expression $\G$ in eq. \eqref{BKC} is given in terms of superspace objects, it is in general non-invariant under all the supersymmetry charges. When it preserves a fraction of supercharges it gives rise to a BPS WS. How many supercharges are preserved by an assigned WS strongly depends on the choice of the supersurface, which eventually translates into the choice of the PCO. For instance, PCO \eqref{BKD} breaks supersymmetry completely, since the corresponding localising condition $\theta^\a \!= \!0 \; \forall \a$ is not invariant under transformations of the form $\theta^\a \to \theta^\a + \e^\a$.   Changing surface $\S \to \S'$ to improve the degree of supersymmetry amounts to changing ${\mathbb Y}^{(0|m)}_\S \to {\mathbb Y}^{(0|m)}_{\S'}$, but as we have already mentioned, the two PCOs necessarily differ by a $d$-exact term (see eq. \eqref{PCOAB}). Therefore, a $d$-varying supersymmetry mechanism can be implemented to span the whole spectrum of BPS WS, as we are going to discuss in the next section.

\section{Super Wilson Surfaces in six dimensions}\label{six} \setcounter{equation}{0}

The previous construction is completely general and can be adapted to different dimensions. In particular, if we fix $n=11$ and $m=32$ in eq. \eqref{AKAF} in principle we obtain a consistent definition of WS in M-theory. 

By dimensional reduction from eleven to six dimensions we land on the $N=(2,0)$ superconformal theory living on one M5-brane. This is a convenient framework where constructing surface operators explicitly. In fact, this is supposed to be a theory of one tensor multiplet which contains a 2-form suitable for defining WS. 

As described in \cite{Howe:1983fr,Bergshoeff:1985mz,Claus:1997cq,Ferrara:2000xg}, the tensor multiplet is given in terms of an anti-symmetric, $\Omega$-traceless\footnote{We refer to appendix \ref{appendix} for notations and conventions of six-dimensional superspace.}  superfield $W^{AB}(x, \theta)$, 
satisfying the superspace constraints and the reality condition
\begin{eqnarray}
\label{TMA}
D^{(A}_\a W^{B) C} =0\,, ~~~~~ \overline W_{AB} = \Omega_{AC} \Omega_{BD} W^{CD}
\end{eqnarray}
Using the algebra of covariant derivatives (\ref{SS6F}), one can show that the superfield has the following
$\theta$-expansion
\begin{eqnarray}
\label{TMB}
W^{[AB]} = \varphi^{[AB]} + \lambda^{[A}_\a \Omega^{B] C}\theta^\a_C + H_{(\a\b)} \theta^{\a[A} \theta^{B] \b} + {\rm derivative~terms}
\end{eqnarray}
where the field components $ \varphi^{[AB]},  \lambda^{A}_\a , H_{(\a\b)}$, which are $5$ scalar fields, 8 fermionic fields and 
3 bosonic fields (self-dual anti-symmetric tensor) are put on-shell
\begin{eqnarray}
\label{TMC}
\partial^2  \varphi^{[AB]} =0 \,, ~~~~~ \partial^{\a\b} \lambda^A_\b =0\,, ~~~~ \partial^{\a\b} H_{\b\gamma} =0
\end{eqnarray}
The latter equation implies that the three form $H_{\mu\nu\rho} \equiv \gamma_{\mu\nu\rho}^{\a\b} H_{\a\b}$ is the curl of a 2-form
\begin{eqnarray}
\label{TMD} 
H_{\mu\nu\rho} = \partial_{[\mu} B_{\nu\rho]} 
\end{eqnarray}
A geometrical formulation in superspace can be obtained by adopting the rheonomic construction. Merging $B_{\mu\nu}$ into the general expansion of a $(2|0)$-form 
\begin{eqnarray}\label{TME2}
B^{(2|0)}= B_{\mu\nu} V^\mu V^\nu +  B_{\mu\a}^A V^\mu \psi^\a_A + B_{\a\b}^{AB} \psi^\a_A \psi^\b_B
\end{eqnarray}
and computing the corresponding curvature $H^{(3|0)} = d B^{(2|0)}$, by imposing conventional constraints  (vanishing of  spinorial components of the curvature) and using Bianchi identities one obtains 
\begin{eqnarray}
\label{TMF}
H^{(3|0)} &=&  V^\mu V^\nu V^\rho \, (\gamma_{\mu\nu\rho})^{\a\b}  D_\a^A D_\b^B W_{AB}  +  V^\mu V^\nu   (\psi_A \gamma_{\mu\nu} D_B) W^{AB} +  V^\mu  (\psi_A \gamma_{\mu} \psi_B) W^{AB} \nonumber \\
&\equiv& V^\mu V^\nu V^\rho \, H_{\mu\nu\rho} + V^\mu V^\nu \psi_A \, H^A_{\mu\nu} + V^\mu  \psi_A \psi_B \, H_\mu^{AB}
\end{eqnarray}
To compute the number of on-shell degrees of freedom one needs to take into account the gauge freedom $\delta B^{(2|0)} = d \Lambda^{(1|0)}$.

Now, using in \eqref{BKC} the 2-form defined in \eqref{TME2} we obtain the supersurface operator for the effective field theory living on the $M5$-brane. These operators can be obtained from their eleven dimensional counterparts by dimensional reduction.

\subsection{Generalized Surface Operators}\label{WMSO}

As discussed in \cite{Cremonini:2020mrk}, in four-dimensional $N=4$ SYM theory it is possible to obtain the generalised Wilson-Maldacena (super)loop, including additional couplings to scalars, from an ordinary (super)Wilson loop by shifting the original PCO by an exact term. Here we investigate whether a similar pattern holds for WS in six dimensions. 

To this end, we first observe that if in the general expression of $\G$ for a superWS in six dimensions
\begin{eqnarray}
\label{M5A}
\Gamma = \int_{{\cal SM} \times \Delta} \!\!\!\! \!\!\!\! B^{(2|0)} \wedge {\mathbb Y}^{(6|16)}_\Sigma
\end{eqnarray} 
we modify the PCO  by the addition of an exact term 
\begin{eqnarray}
\label{M5B}
{\mathbb Y}^{(6|16)}_\Sigma \longrightarrow {\mathbb Y}^{(6|16)}_\Sigma + d \Omega^{(5|16)}
\end{eqnarray}
the resulting operator gets modified as 
\begin{eqnarray}
\label{m5C}
\G  \longrightarrow \Gamma' = \G  + \int_{{\cal SM} \times \Delta}  \!\!\!\! \!\!\!\!  H^{(3|0)} \wedge \Omega^{(5|16)}  
\end{eqnarray}
where the $(3|0)$-superfield strength is given in \eqref{TMF}. The second term originates from integration by parts assuming trivial boundary conditions. 
Now, generalizing what has been done in \cite{Cremonini:2020mrk} for one dimensional contours, we choose $\Omega^{(5|16)}$ to be given by
\begin{eqnarray}
\label{m5D}
\Omega^{(5|16)} &=& 
dz d\bar z  \prod_{\rho=1}^{16}\Big(\theta^\rho - \theta^\rho(z, \bar z)\Big) 
\prod_{\mu=1}^6 \delta\Big(x^\mu - x^\mu (z, \bar z)\Big)  
 \nonumber \\
&~&\qquad \qquad \times \epsilon_{\mu_1 \dots \mu_6} V^{\mu_1} \dots V^{\mu_5} \, N_{AB} (\gamma^{\mu_6})^{[\a\b]} 
\iota^{A}_{~\alpha} \iota^{B}_{~\beta} \delta^{16} (\psi) 
\end{eqnarray}
where $\iota^A_\alpha$ is the contraction respect to the fermionic vector field $D^A_\a$, and $N_{AB}$ 
is a pseudoreal, antisymmetric\footnote{The expression $(\gamma^{\mu_6})^{[\a\b]} 
\iota^{A}_{~\alpha} \iota^{B}_{~\beta}$ is anti-symmetric in $A,B$ since $\iota^A_\alpha$ and $\iota^B_\beta$
commute between them.} tensor of the $USp(4)$ R-symmetry group, satisfying 
$\overline N_{AB} = \epsilon_{ABCD} N^{CD}$.   

Inserting this particular $\Omega^{(5|16)} $ form in (\ref{m5C}) and taking into account that $H^{(3|0)}$ is the sum of three pieces with 
decreasing powers of $V$'s, it is easy to realise that only the term linear in $V$ in (\ref{TMF}) survives. Adapting the expression in \eqref{BKC} for $\G$ to six dimensions 
and combining it with $\int H^{(3|0)} \wedge \Omega^{(5|16)} $ as in \eqref{m5C} we are finally led to
\begin{eqnarray}
\label{m5E}
\hspace{-0.9cm}\G' = \int_\Sigma   \Big(B_{\mu\nu} \Pi^\mu_z \Pi^\nu_{\bar z} \! + \! 
B_{\mu\a}^A (\Pi^\mu_z \, \partial_{\bar z}\theta^\a_A + \Pi^\mu_{\bar z} \, \partial_{z}\theta^\a_A) + 
B_{\a\b}^{AB} \, \partial_{z}\theta^\a_A \, \partial_{\bar z}\theta^\beta_B + N_{AB} W^{AB}\Big) dz d\bar z
\end{eqnarray}
where $W^{AB}$ is the tensor superfield \eqref{TMB}, $B_{\mu\nu}, B_{\mu\a}^A, B_{\a\b}^{AB}$ are the components of the corresponding $(2|0)$-form 
given in \eqref{TME2} and $\Pi^\mu_i = \partial_i x^\mu + \theta_A \Omega^{AB} \g^\mu \partial_i \theta_B$ for $i=z, \bar z$. We note that the last term of $\G'$ has an additional symmetry. In fact, as a consequence of the $\Omega$-traceless property of $W^{AB}$, it is invariant under 
$\delta N_{AB} = N \Omega_{AB}$. This symmetry is useful to remove one degree of freedom from the $N_{AB}$ tensor. 

Equation \eqref{m5E} is the natural definition of a generalized superWS in superspace. Its lowest component, obtained by setting $\theta^\a =0$ everywhere, coincides with the operator introduced in \cite{Gustavsson:2004gj} and more recently studied in \cite{Mezei:2018url,Drukker:2020dcz,Drukker:2020swu}, which includes couplings to the five scalar fields $\varphi^{[AB]}$ of the tensor multiplet, in analogy with the Wilson-Maldacena loop.

\subsection{BPS Surface Operators}\label{susy}

We now study under which conditions a (super)WS preserves a certain amount of supersymmetry. This amounts to determine and solve the Killing spinor equations for the assigned operator. 

We will consider the generic operator
\begin{eqnarray}\label{general}
\Gamma_\zeta [\S] = \int_{{\cal SM} \times \Delta} \!\!\!\! \!\!\!\! B^{(2|0)} \wedge \left( {\mathbb Y}^{(6|16)}_\Sigma + \zeta \, d \Omega^{(5|16)} \right)
\end{eqnarray} 
which interpolates between the WS in \eqref{M5A} (for $\zeta =0$) and the generalized one \eqref{m5C} (for $\zeta = 1$), with $\Omega^{(5|16)}$ given in \eqref{m5D}. 

Expression \eqref{general} is manifestly invariant under superdiffeomorphisms, being the integral of a top form on the entire (extended) supermanifold. Since on superforms and PCOs an infinitesimal superdiffeomorphism generated by a vector field $X$ acts as the Lie derivative, $\d_X = \{d, \iota_X\}$,
where $\iota_X$ is the contraction along $X$, the invariance of $\G$ can be explicitly written as
\begin{equation}\label{deltag}
\d_X \G_\zeta [\S] = \int_{{\cal SM} \times \Delta} \left[ \iota_X H^{(3|0)} \! \wedge \! \left( {\mathbb Y}^{(6|16)}_\Sigma + \zeta \, d \Omega^{(5|16)} \right) + H^{(3|0)} \!\wedge \! \iota_X \! \left( {\mathbb Y}^{(6|16)}_\Sigma + \zeta \, d \Omega^{(5|16)} \right)\right] \equiv 0
\end{equation} 
Here we have used the $d$-closure of the PCO and assumed the absence of boundary contributions. The first term of the integrand corresponds to the variation in form of $\G_\zeta$, whereas the second term, being associated to the variation of the PCO, arises from the variation of the supersurface. This identity thus states that the variation in form of $\G_\zeta$ induced by the $X$-transformation is always compensated by the variation of the supersurface $\S$. In particular, this implies that the $X$-diffeomorphism is a symmetry for $W_S[\S,\zeta] = e^{\G_\zeta[\S]}$ if it leaves the supersurface invariant, $\d_X \S=0$. Differently stated, the set of WS invariances coincides with the set of $\S$ symmetries. 

A supersymmetry transformation is a particular superdiffeomorphism corresponding to $X \equiv \e = \epsilon^\alpha_A Q_\alpha^A $, with $Q_\a^A$ being the supersymmetry charges defined in \eqref{SS6H}. Applying the previous reasoning we can trade the supersymmetry invariance equation $\d_\e W_S[\S,\zeta]=0$ with the condition $\d_\e \S=0$. This is indeed the Killing spinor equation which can be used to classify BPS WS. 

Explicitly, from eq. \eqref{deltag} this equation reads
\begin{equation}\label{SSA}
	H^{(3|0)} \wedge \iota_\epsilon \left( \mathbb{Y}_\Sigma^{(6|16)} + \zeta \, d \Omega^{(5|16)} \right)\sim 0
\end{equation}
where ``$\sim$'' means that this quantity has to be zero, up to $d$-exact terms.

For simplicity, we begin investigating the $\zeta=0$ case. 
Using the action of $\iota_\epsilon$ on the supervielbein
\begin{equation}\label{SSB}
\iota_\epsilon V^\mu = 2 \epsilon \gamma^\mu \theta, \qquad \iota_\epsilon \psi^\alpha_A = \epsilon^\alpha_A
\end{equation}
the application of $\iota_\epsilon$ to the PCO in \eqref{BKA3} leads to
\begin{eqnarray}\label{SSC} 
&& \iota_\epsilon \mathbb{Y}_\Sigma^{(6|16)} = \\
&& \delta^{(6)} (x  -  x(z, \bar{z})) \, 2 \epsilon \gamma^\mu \theta \iota_\mu \, ( V - \Pi_z dz - \Pi_{\bar{z}} d \bar{z})^6 \, (\theta -  \theta(z, \bar{z}) )^{16} \, \delta^{(16)}(\psi  -  \partial_z{\theta} dz \! - \! \partial_{\bar{z}} \theta d \bar{z} ) \nonumber  \\
&& + \delta^{6} (x - x(z, \bar{z})) \, ( V - \Pi_z dz - \Pi_{\bar{z}} d \bar{z})^6 \, (\theta - \theta(z, \bar{z}) )^{16} \, \epsilon_A \iota^A \, \delta^{(16)}(\psi - \partial_z{\theta} dz - \partial_{\bar{z}} \theta d \bar{z} ) \nonumber 
\end{eqnarray}
Now, multiplying this expression by the $H^{(3|0)}$ expansion in \eqref{TMF}, it is easy to see that the first line in \eqref{SSC} let all the terms in \eqref{TMF} survive, whereas the second line kills all the terms except for the $VV\psi$ and $V \psi \psi$ ones. Assembling everything together, we obtain
\begin{equation}\label{SSD}
	\Big( 4 \epsilon \gamma^\mu \theta \Pi^\nu_z \Pi^\rho_{\bar{z}} H_{\mu \nu \rho} + 2 \epsilon \gamma^\mu \theta \left( \Pi^\nu_z \partial_{\bar{z}} \theta_A - \Pi^\nu_{\bar{z}} \partial_z \theta_A \right) H_{\mu \nu}^A + 8 \epsilon \gamma^\mu \theta \partial_z \theta_A \partial_{\bar{z}} \theta_B H_\mu^{AB}   $$
	$$ \left. - 2 \Pi^\mu_z \Pi^\nu_{\bar{z}} \epsilon_A H_{\mu \nu}^A + 2 \epsilon_A \left( \Pi^\mu_z \partial_{\bar{z}} \theta_B - \Pi^\mu_{\bar{z}} \partial_z \theta_B \right) H_\mu^{AB} \Big) \right|_{\Sigma}  \times {\rm Vol} = 0
\end{equation}
where we have defined
\begin{equation}\label{vol}
{\rm Vol} = \d^{(6)}(x - x(z,\bar{z})) \, V^6 dz d\bar{z} \, (\theta - \theta(z,\bar{z}))^{16} \, \d^{(16)}(\psi)
\end{equation}

This is the most general Killing spinor equation which in principle allows to classify all the BPS supersurfaces in superspace. Its systematic investigation is beyond the scopes of the present paper and is left for the future. Here we consider only the special class of purely bosonic surfaces, namely we set $\theta^\a \left( z , \bar{z} \right) = 0$. In this case the previous equation greatly simplifies and reduces to
\begin{equation}\label{SSE}
	\left. \left( \Pi^\mu_z \Pi^\nu_{\bar{z}} \epsilon_A H_{\mu \nu}^A \right) \right|_{\Sigma}  \times {\rm Vol} = 0 \quad \Rightarrow  \quad \partial_z x^\mu \partial_{\bar{z}} x^\nu  (\epsilon_A   \gamma_{\mu\nu} D_B) W^{AB}  = 0
\end{equation}
where in the last expression all the functions are localized on $\S$ and the $\e_A$ spinor is in general a local function of the point on the surface. If we require this equation to be valid for any $W^{AB}$, the Killing spinor equation that we have to solve is
\begin{equation}\label{SSF}
	\epsilon_A  \, \partial_z x^\mu \partial_{\bar{z}} x^\nu  \gamma_{\mu\nu}    = 0 
\end{equation}

We look for constant $\epsilon_A$ solutions, then corresponding to supersymmetry globally realized on the surface. Non-trivial solutions exist if the $4 \times 4$ matrix $M \equiv \partial_z x^\mu \partial_{\bar{z}} x^\nu  \gamma_{\mu\nu}$ has a non-trivial kernel or, equivalently, if ${\rm det} M =0$. In particular, the rank of the matrix will determine the BPS degree of the corresponding surface operator. 

In order to study this equation in general, it is convenient to trade $M$ for $M^2$ and look for solutions of $ {\rm det} M^2=0$.  In fact, rewriting $M$ as 
\begin{eqnarray}
\label{SSG}
M = \epsilon^{ij} \partial_i x^\mu \partial_j x^\nu  \gamma_{\mu\nu}  = \epsilon^{ij} (\partial_i x^\mu \gamma_{\mu})(\partial_j x^\nu \bar\gamma_{\nu})  
, \qquad i,j = z , \bar z
\end{eqnarray}
and making use of the Clifford algebra and Schouten's identity for the $\e^{ij}$ tensor, its square turns out to be proportional to the  $4 \times 4$ identity matrix
\begin{eqnarray}
\label{SSH}
M^2
&=&2 \left(  (\partial^i x^\mu \partial_i x^\nu \eta_{\mu\nu})^2 - (\partial_i x^\mu \partial_k x^\nu \eta_{\mu\nu})(\partial^i x^\rho \partial^k x^\sigma \eta_{\rho\sigma})  
 \right)  \nonumber \\
 &=& 4 \det 
\left(
\begin{array}{ccc}
  \partial_z x^\mu \partial_z x^\nu \eta_{\mu\nu} \; &   \;  \partial_z x^\mu \partial_{\bar z} x^\nu \eta_{\mu\nu} \\
  \partial_z x^\mu \partial_{\bar z} x^\nu \eta_{\mu\nu}  \;  & \;    \partial_{\bar z} x^\mu \partial_{\bar z} x^\nu \eta_{\mu\nu}  
\end{array}
\right) \times {\mathcal I}
\end{eqnarray}
Therefore, ${\rm det}M^2$ is proportional to the determinant in \eqref{SSH} and it vanishes if  the following equation 
\begin{eqnarray}
\label{SSK}
(\partial_z x^\mu \partial_z x^\nu \eta_{\mu\nu})   (\partial_{\bar z} x^\mu \partial_{\bar z} x^\nu \eta_{\mu\nu}) -  (\partial_z x^\mu \partial_{\bar z} x^\nu \eta_{\mu\nu})^2=0
\end{eqnarray}
is satisfied. This is a non-trivial equation for the $x^\mu$ coordinates of the surface and selects a subset of BPS surfaces.

To solve equation \eqref{SSK} we embed the two-dimensional surface into a three-dimensional manifold ${\cal N} \subset {\cal M}$ where ${\cal M}$ is the six-dimensional Minkowskian bosonic slice of the supermanifold with signature $(-, +, \dots, +)$.  

We begin by considering a timelike three-dimensional slice. In order to prove that at least one non-trivial solution of \eqref{SSK} exists, we make the easiest ansatz  
\begin{eqnarray}
\label{SSL}
&&x^\mu(z,{\bar z}) = (f(z, {\bar z}), 0, 0, 0, z, {\bar z})
\end{eqnarray}
where $f$ is a smooth function to be determined. Equation \eqref{SSK} then reduces to the well-known Light Ray Partial Differential Equation (see for instance \cite{DiffEquat})
\begin{eqnarray}
\label{SSM}
 (\partial_z f)^2 + (\partial_{\bar z} f)^2 = 1 
\end{eqnarray}
Using an adapted $\g$-matrix representation (see appendix \ref{appendix}) the corresponding $M$ matrix takes the $2 \times 2$ block form
\begin{eqnarray}
M = \left(
\begin{array}{ccc}
 {\mathbb A} \; &   \;  \; \; 0 \\
  0  \;  & \;    -{\mathbb A }
\end{array}
\right) \, , \qquad \qquad \quad {\mathbb A} = -(\partial_z f) \s_2 + (\partial_{\bar z} f) \s_3 - i \s_1
\end{eqnarray}
where eq. \eqref{SSM} ensures ${\rm det} {\mathbb A} = 0$ and necessarily corresponds to an even number of zero eigenvalues for $M$. Therefore, excluding the case of a null matrix, we conclude that any solution to equation \eqref{SSM} provides a rank-2 matrix $M$ and yields a 1/2 BPS WS.

One class of 1/2 BPS solutions is given by linear functions of the form
\begin{eqnarray}
\label{SSN}
f(z,{\bar z}) = C_1 z + C_2 {\bar z} + C_3\,, ~~~~~ C_1^2 + C_2^2 =1
\end{eqnarray}
For fixed $C_1, C_2, C_3$ constants, it describes a plane immersed in three dimensions with one time direction. 
Another class of 1/2 BPS solutions encodes quadratic functions of the form
\begin{equation}\label{SSO}
	f^2 \left( z , \bar{z} \right) = \left( z - C_1 \right)^2 + \left( \bar{z} - C_2 \right)^2
\end{equation}
which for fixed constants describes a spherical two-dimensional wavefront.

Things drastically change  if we consider immersion into a spacelike three dimensional submanifold. This amounts to modify ansatz \eqref{SSL} for instance as
\begin{eqnarray}
\label{SSL2}
x^\mu(z,{\bar z}) = (0, 0, 0, z, {\bar z},f(z, {\bar z}))
\end{eqnarray}
As a consequence of the change in signature, it is easy to realize that constraint \eqref{SSM} gets substituted by 
\begin{eqnarray}
\label{SSM2}
 (\partial_z f)^2 + (\partial_{\bar z} f)^2 = -1 
\end{eqnarray}
and does not allow for any real solution. Therefore, we conclude that in Minkowski signature there are no spacelike 1/2 BPS WS. This result resembles the Wilson loop situation, where no spacelike BPS Wilson operators exist in Minkowski spacetime \cite{Ouyang:2015ada}. 

 \vskip 10pt
We now study the BPS constraint  \eqref{SSA} in the generalized case, $\zeta \neq 0$. This requires evaluating also the second term $H^{(3|0)} \wedge \iota_\epsilon  d \Omega^{(5|16)} $. Since from eq. \eqref{m5D} we easily obtain\footnote{For avoiding cluttering we neglect $(z,\bar{z})$ indices of the $N_{AB}$ components.}
\begin{eqnarray}
	\label{SSWMB} d \Omega^{(5|16)} &=& \partial_\mu \left[ dz d \bar{z} \left( \theta - \theta ( z, \bar{z} ) \right)^{16} \delta^{(6)} \left( x - x ( z , \bar{z} ) \right) V^6 N_{AB} \gamma^\mu \iota^A \iota^B \delta^{(16)} \left( \psi \right) \right] + \\
	\nonumber && \hspace{-0.2cm} - 2 D^A \left[ dz d \bar{z} \left( \theta - \theta ( z, \bar{z} ) \right)^{16} \delta^{(6)} \left( x - x ( z , \bar{z} ) \right) \epsilon_{\mu_1 \ldots \mu_6} V^{\mu_1} \ldots V^{\mu_5} N_{AB} \gamma^{\mu_6} \iota^B \delta^{(16)} \left( \psi \right) \right]
\end{eqnarray}
the contraction $\iota_\epsilon$ gives rise to
\begin{eqnarray}\label{SSWMC}
&& \hspace{-0,8cm} \iota_{\epsilon} d \Omega^{(5|16)} = 2 \epsilon \gamma^\nu \theta \partial_\mu \left[ dz d \bar{z} \left( \theta - \theta ( z, \bar{z} ) \right)^{16} \delta^{(6)} \left( x - x ( z , \bar{z} ) \right) \iota_\nu V^6 N_{AB} \gamma^\mu \iota^A \iota^B \delta^{(16)} \left( \psi \right)  \right] \\
\nonumber &+& \epsilon_C \partial_\mu \left[dz d \bar{z} \left( \theta - \theta ( z, \bar{z} ) \right)^{16} \delta^{(6)} \left( x - x ( z , \bar{z} ) \right) V^6 N_{AB} \gamma^\mu \iota^A \iota^B \iota^C \delta^{(16)} \left( \psi \right) \right] \\
\nonumber &+& 4 \epsilon \gamma^\nu \theta D^A \left[ dz d \bar{z} \left( \theta - \theta ( z, \bar{z} ) \right)^{16} \delta^{(6)} \left( x - x ( z , \bar{z} ) \right) \iota_\nu \epsilon_{\mu_1 \ldots \mu_6} V^{\mu_1} \ldots V^{\mu_5} N_{AB} \gamma^{\mu_6} \iota^B \delta^{(16)} \left( \psi \right) \right] \\
\nonumber &-& 2 \epsilon_C D^A \left[ dz d \bar{z} \left( \theta - \theta ( z, \bar{z} ) \right)^{16} \delta^{(6)} \left( x - x ( z , \bar{z} ) \right) \epsilon_{\mu_1 \ldots \mu_6} V^{\mu_1} \ldots V^{\mu_5} N_{AB} \gamma^{\mu_6} \iota^B \iota^C \delta^{(16)} \left( \psi \right) \right]
\end{eqnarray}
This result, when multiplied by $H^{(3|0)}$ in \eqref{TMF}, leads to
\begin{equation}
	\left. \Big( - 4 \partial_\mu H_{\nu}^{AB} \gamma^\mu N_{AB} \epsilon \gamma^\nu \theta - 4 D^A \gamma^\mu H^B_{\nu \mu} \epsilon \gamma^\nu \theta N_{AB} + 4 D^A H_\mu^{BC} \epsilon_C N_{AB} \gamma^\mu \Big) \right|_{\Sigma} \times \text{Vol}
\end{equation}
where the volume form is given in \eqref{vol}. Summing this result with \eqref{SSD} we obtain the generalized Killing spinor equations in superspace. 

As before, the discussion simplifies in the particular case  $\theta \left( z , \bar{z} \right) = 0$, that is when we look for ordinary BPS surfaces. In fact, from the previous result we simply obtain
\begin{equation}\label{SSWMD}
	H^{(3|0)} \wedge \iota_\epsilon  d \Omega^{(5|16)}  = 4 D^A \gamma_\mu W^{BC} \epsilon_C N_{AB} \gamma^\mu \, \times \, \text{Vol}
\end{equation}
Here the numerical coefficient comes from manipulating gamma matrices and the superderivative has been moved to act on $W$ by using the Leibniz rule. Combining this result with \eqref{SSE} and paying attention to the relative coefficients, we obtain  
\begin{equation}\label{SSWME}
	\left( \partial_z x^\mu \partial_{\bar{z}} x^\nu \Omega_{BC} \epsilon_A \gamma_{\mu \nu} - 12\zeta \, N_{BC} \epsilon_A \right) D^B W^{AC} = 0 
\end{equation}
By suitably rescaling $N_{BC}$ and requiring this equation to be valid for any $W^{AC}$ configuration we finally land on
\begin{equation}\label{SSWMF}
	\epsilon_A  \, \left( \partial_z x^\mu \partial_{\bar{z}} x^\nu \Omega_{BC} \gamma_{\mu \nu} - \zeta \, N_{BC}  \right) = 0 
\end{equation}
For $\zeta = 1$ this  coincides with the Killing spinor equation discussed in \cite{Mezei:2018url,Drukker:2020dcz}. Non-vanishing solutions require the following consistency condition to be valid
\begin{eqnarray}
\label{cirpA2}
\det{\Big(\Pi_i^\mu \Pi_j^\nu \eta_{\mu\nu}\Big) - N_{A}^B N^{A}_B } = 0
\end{eqnarray}

\section{Kappa Symmetry} \setcounter{equation}{0}

We now study the behavior of superWS under \emph{kappa symmetry}, that is under transformations generated by the vector field $\tilde{\kappa} = \kappa^\alpha_A D^A_\alpha$ with supercovariant derivatives given in \eqref{SS6F}. In the present section we will restrict to six dimensions\footnote{Kappa-symmetry transformations for the $(6|16)$-dimensional supermanifold are given in \eqref{Ktransfs}.}, for which we have the general decomposition of the superform $H^{(3|0)}$, eq. \eqref{TMF}. However, the results that we find do not rely on this particular choice and can be easily adapted to other dimensions. 

According to the general discussion above, the generic operator \eqref{general} is invariant when the following condition is satisfied
\begin{equation}\label{KSA}
	H^{(3|0)} \wedge \iota_{\tilde{\kappa}} \left( \mathbb{Y}_\Sigma^{(6|16)} + \zeta \, d \Omega^{(5|16)} \right) \sim 0
\end{equation}

We first study the $\zeta =0$ case. Recalling the action of kappa symmetry on the six dimensional supervielbeins, eqs. \eqref{KSB}, we can easily compute 
\begin{equation}\label{KSC}
	\iota_{\tilde{\kappa}} \mathbb{Y}_{\Sigma}^{(6|16)} = \d^{(6)} (x - x(z, \bar z)) \, ( V - \Pi_z dz - \Pi_{\bar z} d\bar z)^6 \, \wedge \, (\theta - \theta(z, \bar z) )^{16} \, \kappa^\alpha_A \iota^A_\alpha \d^{(16)}(\psi - \partial_z{\theta} dz - \partial_{\bar z} \theta d\bar z )
\end{equation}
It follows that contracting with $H^{(3|0)}$ the only non-zero terms come from the $VV\psi$ and $V\psi\psi$ terms of \eqref{TMF}. Therefore, we  obtain
\begin{equation}\label{KSDA}
	\epsilon^{ij} \, \Pi^\mu_i \Pi^\nu_j \Omega_{BC}  \left( D^B W^{AC} \right) \gamma_{\mu \nu} \kappa_A + 
	\epsilon^{ij}  \, \Pi^\mu_i \partial_{j} \theta_A  \gamma_{\mu} \kappa_B  W^{AB}  
	= 0 \; , \qquad i,j = z, \bar{z}
\end{equation}
where $\Pi^\mu_i = \partial_i x^\mu + \theta_A \Omega^{AB} \g^\mu \partial_i \theta_B$. Since we require this equation to be true independently of the particular values of $W^{AB}$, the two terms have to vanish separately. In order to study these two conditions we make the conventional ansatz $\kappa_A = \e^{ij} \gamma_{\mu \nu} \left( \Pi^{\mu}_i \Pi^{\nu}_j \right) K_A$ and look for constant $K_A$ solutions in various examples, with an increasing level of generality. 

As the simplest case, we look for solutions in the subset of ordinary surfaces, that is we set  $\theta_A = 0$. 
Following a procedure similar to the one that in the case of supersymmetry led to \eqref{SSH}, we  obtain that non-vanishing constant $K_A$ solutions  exist if the supersurface coordinates satisfy the following condition 
\begin{eqnarray}
\label{kaC}
 {\rm det}(\Pi^\mu_i \Pi^\nu_j \eta_{\mu\nu})=0 \, , \qquad \quad i,j = z, \bar{z}
\end{eqnarray}
This condition has an interesting interpretation from the point of view of the dual geometry. In the AdS$_7$/CFT$_6$ correspondence a surface operator $W[\S]$ for the $N=(2,0)$ superconformal field theory (SCFT) living on a M5-brane is holographically dual to an extremized supermembrane worldvolume whose boundary coincides with the $\S$ surface on the M5-brane \cite{Maldacena:1998im}. If we consider the standard action of a supermembrane in eleven dimensional notation as given in \cite{Berkovits:2002uc}, the equations of motion for the worldvolume metric lead to the worldvolume reparametrization constraints
\begin{eqnarray}
\label{kaD}
P^2 + {\rm det}(\Pi^M_i \Pi^N_j \eta_{MN}) = 0 \; , \qquad \quad P_M  \Pi^M_i  = 0
\end{eqnarray}
where $P^M$ is the momentum of the membrane\footnote{Obtained by taking the derivative of the Lagrangian with respect to the time derivative of the 11-dimensional 
coordinates $\partial_0 x^\mu$.} while $\Pi^M_i$ are the spatial (super)tangent vectors to the membrane. 
These constraints ensure that the M2-brane action is invariant under kappa-symmetry transformations \cite{Berkovits:2002uc}. In particular, kappa-symmetry transformations for the M2-brane supercoordinates read
\begin{eqnarray}
\label{kaA}
\delta \theta = (\Gamma_M P^M + \frac12 \epsilon^{ij} \Gamma_{MN} \Pi^M_i \Pi^N_j) K
\end{eqnarray}
for some spacetime spinor $K$. It is easy to see that for a static supermembrane, that is setting $P^M =0$, these transformations coincide with the ones that we used, $\d_{\tilde \kappa} \theta_A = \e^{ij} \gamma_{\mu \nu} \left( \Pi^{\mu}_i \Pi^{\nu}_j \right) K_A$ and the equations of motion \eqref{kaD} are nothing but constraint \eqref{kaC} for kappa-symmetry invariance of the WS. 
Therefore, this constraint  can be interpreted as the requirement for the static membrane to be kappa symmetric. Since for $\theta_A = 0$ the tangent vectors reduce to
$\Pi^\mu_i = \partial_i x^\mu$, remarkably the kappa-symmetry constraint coincides with the constraint for supersymmetry studied above. 

Now, we look for more general solutions with $\partial_j \theta_A \neq 0$. In this case also the second term in eq. (\ref{KSDA}) gives a non-trivial constraint for kappa-symmetry invariance. Inserting there $\kappa_A = \e^{ij} \gamma_{\mu \nu} \left( \Pi^{\mu}_i \Pi^{\nu}_j \right) K_A$, using the Schouten's identities and Clifford algebra rules it can be cast in the following form
\begin{eqnarray}
\label{WEA}
\Big(\delta^{ij} \Pi^\mu_i \Pi^\nu_j \eta_{\mu\nu} \Pi^\rho_k \delta^{kl} -  
\delta^{ij} \Pi^\mu_i \Pi^\nu_k \eta_{\mu\nu} \Pi^\rho_j \delta^{kl}\Big ) \partial_l\theta_{[A} \gamma_\rho K_{B]}  =0
\end{eqnarray}
We introduce the matrix $G_{ij} = \Pi^\mu_i \Pi^\nu_j \eta_{\mu\nu}$ that satisfies $\det(G_{ij})=0$, as expressed by (\ref{kaC}). In terms of $G$ equation (\ref{WEA}) reads
\begin{eqnarray}
\label{WEB}
  \Pi^\rho_k \Big( \delta^{k}_l -  
\frac{G_{l}^{k}}{\tr(G)} \Big )\partial^l\theta_{[A} \gamma_\rho K_{B]}  =0
\end{eqnarray}
where $\tr(G) = \delta^{ij} G_{ij}$. Using the identity 
$G_{i}^{~j} G_{j}^{~k} = G_{i}^{~k} \tr(G) - \delta_i^{~k} \det(G_{ij})$
it is easy to realise that constraint \eqref{kaC} implies that the matrix $\Big( \delta_l^{k} - \frac{G_{l}^{k}}{\tr(G)}\Big )$ is a projector. 
It follows that equation (\ref{KSDA}) admits further solutions when $\partial_i \theta^\a_A$ is in 
the kernel of this projector. We note that this is 
the usual framework of kappa-symmetric dynamics: The equations of motion for the fermionic coordinates 
are wave equations with a degenerate wave operator.

\subsection{Kappa Symmetry for generalized Wilson Surfaces}

We now study the $\zeta \neq 0$ case corresponding to a generalized WS. Since the first term in \eqref{SSA} has been already discussed above we focus only on the 
$\zeta$-term. 

Applying the $\iota_{\tilde{\kappa}}$ operator to \eqref{SSWMB}, the first term leads to an expression proportional to $\displaystyle \iota^3 \delta^{16} \left( \psi \right) $. Since in \eqref{TMF} the term proportional to $\psi^3$ is zero, it follows that the only non-trivial expression comes from the second term of \eqref{SSWMB}, and we obtain 
\begin{eqnarray} \label{KSWMC} 
&& \hspace{-0.2cm}  H^{(3|0)} \wedge \iota_{\tilde{\kappa}} d \Omega^{(5|16)} = \\
	\nonumber &&  \hspace{-0.2cm} V^\mu \psi_A \gamma_\mu \psi_B W^{AB} dz d \bar{z} \delta^6 \left( x \! - \! x \left( z , \bar{z} \right) \right) \kappa_C \iota^C \left[ \psi_D D^D \Big( \left( \theta \! - \! \theta \left( z , \bar{z} \right) \right)^{16} \iota_\nu V^6 N_{EF} \gamma^\nu \iota^E \iota^F \delta^{16} \left( \psi \right) \Big) \right]
\end{eqnarray}
We can now move the spinorial derivative on $W^{AB}$ and perform all the contractions to obtain
\begin{equation}
	24 \, N_{AC} \kappa_B^\alpha D^A_\alpha W^{BC}  \times \text{Vol} 
\end{equation}
where Vol has been defined in \eqref{vol}. 
Inserting this expression in \eqref{KSA} and combining with the rest of the terms (see eq. \eqref{KSDA}) we finally obtain
\begin{equation}\label{cirpB}
	\left( \Pi^\mu_z \Pi_{\bar{z}}^\nu \Omega_{BC} \kappa_A \gamma_{\mu \nu} - 12\zeta \, N_{BC} \kappa_A \right) D^B W^{AC} - \gamma_\mu W^{AB} \left( - \Pi_z^\mu \partial_{\bar{z}} \theta_A + \Pi^\mu_{\bar{z}} \partial_z \theta_A \right) \kappa_B = 0 
\end{equation}
As before, if we require this equation to be satisfied for any $W^{AB}$ the two terms have to vanish separately. 
In order to solve these two equations we make the more general ansatz $\kappa_A = \epsilon^{ij} (\g_{\mu \nu} \Pi_i^\mu \Pi^\nu_j \delta_A^{~B} + N_{ij, A}^{~~B}) K_B$. Considering for instance the first bracket in \eqref{cirpB}, suitably rescaling $N_{AB}$ we obtain
\begin{eqnarray}
\label{cirpA}
\det{\Big(\Pi_i^\mu \Pi_j^\nu \eta_{\mu\nu}\Big) - N_{A}^B N^{A}_B } = 0
\end{eqnarray}
As for the case of Wilson-Maldacena loops \cite{Maldacena:1998im}, the extra terms proportional to the $N_{AB}$ scalar couplings arise from the dimensional reduction to six dimensions of the eleven-dimensional constraint $\det{\Big(\Pi_i^M \Pi_j^N \eta_{MN}\Big) } = 0$ for the static supermembrane (see eq. \eqref{kaD}). 
Remarkably, this constraint coincides with \eqref{cirpA2}) which ensures supersymmetry invariance. Therefore, kappa-symmetry in eleven dimensions implies BPS properties in six dimensions. 

The second piece of eq. (\ref{cirpB}) can be analyzed along the same lines as above.

\section{Tensor Currents} \setcounter{equation}{0} \label{currents}

The geometric construction of (super)surface operators given in sections \ref{sect:bosonicWS} and \ref{sect:susyWS} can be easily generalized to define (super)hypersurface operators generated by a $(p|0)$-form. In a $(n|m)$-dimensional supermanifold ${\cal SM}$, definition \eqref{AKAF} generalizes to
\begin{eqnarray}
\label{AKAFgen}
W_p[\Sigma] = e^{ \Gamma } \,, ~~~~~  \Gamma = \int_{\cal SM} \!\!\!\!  B^{(p|0)} \wedge {\mathbb Y}^{(n-p|m)}_\Sigma
\end{eqnarray} 
where now $\S$ is a hypersurface of dimensions $(p|0)$. Setting the Grassmann coordinates to zero, this equation is also a generalization of the WS in \eqref{WSAbis}.

Surface operators and, more generally, higher dimensional hypersurface operators describe objects charged under generalized global symmetries generated by tensor currents \cite{Gaiotto:2014kfa}. In order to embed this relation within our geometrical approach, in this section we formulate tensorial conservation laws in curved (super)manifolds using the PCO formalism. The main goal is to generalize the construction of \cite{Gaiotto:2014kfa} and define conservation laws in superspace. Moreover, we investigate general conditions which allow to span the whole set of conserved charges, both for tensor currents and supercurrents, and find the corresponding charged objects. 

Following the recent classification of \cite{Gaiotto:2014kfa,Cordova:2018cvg,Seiberg:2019vrp} we first investigate the case of $U(1)$ $p$-tensor symmetries. In section \ref{supercurrents} we then construct the supersymmetric version of tensorial conservation laws and interpret the super-hypersurface operators, in particular the superWS introduced in the previous sections, as the corresponding charged objects.

As a warming-up, we first review in geometrical language the case of an ordinary bosonic vector current $J^\mu = (J^0, J^i)$ in $n$-dimensions, whose conservation law in Minkowski signature reads
\begin{eqnarray}
\label{TCA}
\partial_\mu J^\mu = 0 \qquad \Leftrightarrow \qquad \partial_0 J_0 = \partial_i J^i
\end{eqnarray}
Accordingly, we foliate the spacetime manifold as ${\cal M}^{(n)} = {\cal M}^{(n-1)} \times I$ where $I$ is an open time interval. We endow the space-slice $ {\cal M}^{(n-1)}$ with a metric structure $g = g_{ij} dx^i \otimes dx^j$ and denote by $\star$ the Hodge dual on $ {\cal M}^{(n-1)}$ with respect to $g$. The conservation law (\ref{TCA}) can then be rephrased as follow 
\begin{eqnarray}
\label{TCB}
\partial_0 J_0 = \star d \star J^{(1)} \equiv d^\dagger J^{(1)} 
\end{eqnarray}
where $J^{(1)}$ is the 1-form on $ {\cal M}^{(n-1)}$ and $d$ is the spatial differential. The corresponding conserved charge is given by  
\begin{eqnarray}
\label{TCC2}
Q = \int_ {{\cal M}^{(n-1)}} \star J_0
\end{eqnarray}
and thanks to the conservation law in \eqref{TCB}, is trivially conserved
\begin{eqnarray}
\label{TCC}
 \partial_0 Q = \int_{{\cal M}^{(n-1)}}\partial_0( \star J_0) =  \int_ {{\cal M}^{(n-1)}} d (\star J^{(1)}) = 0
\end{eqnarray}
as long as non-trivial boundary terms are absent. 

In principle the conserved charge could be rewritten as an integral of a top form on the entire manifold  
\begin{eqnarray}
\label{TCDt}
Q = \int_{{\cal M}^{(n)}} (\star J_0) \wedge  \mathbb{Y}^{(1)}\, ,  \;  \qquad  \qquad 
\mathbb{Y}^{(1)} = \d(x^0) dx^0 = \hat d \, \Theta(x^0)
\end{eqnarray}
where $\mathbb{Y}^{(1)}$ is the PCO that localizes the integral in the time direction. Here $\hat d$ indicates the differential on the entire manifold ${\cal M}^{(n)}$. We note that the relation $\mathbb{Y}^{(1)} = \hat d \, \Theta(x^0)$ does not contradict the general statement that PCOs are closed but not exact, since we have enlarged the domain to distributions with non-compact support. 

Keeping this in mind, in the rest of the discussion we will restrict all the integrations to the constant time slice ${\cal M}^{(n-1)}$, so avoiding the use of $\mathbb{Y}^{(1)} $.  This PCO can be easily reinserted whenever it is more convenient to write $Q$ as the integral of a spacetime top form.

\subsection{$(p+1)$-form Currents}

The generalization of conservation law \eqref{TCA} to tensorial currents has been first discussed in \cite{Gaiotto:2014kfa,Cordova:2018cvg,Seiberg:2019vrp}. 
Here we consider the case of a $U(1)$ $(p+1)$-form current decomposed as $\hat J^{(p+1)} = ( J_0^{(p)},  J^{(p+1)})$, where 
$J_0^{(p)}$ and $J^{(p+1)}$ are $p$ and $(p+1)$-forms in the space-slice ${{\cal M}^{(n-1)}}$, respectively. The spacetime conservation law for the $\hat J^{(p+1)}$  current can be expressed in terms of the following two equations
\begin{eqnarray}
\label{cicA}
\partial_0 J^{(p)}_0 = d^\dagger  J^{(p+1)}\,, ~~~~~   d^\dagger   J^{(p)}_0 =0 
\end{eqnarray}
or equivalently of their Hodge duals
\begin{eqnarray}
\label{cicB}
\partial_0 \star   J^{(p)}_0 = d \star   J^{(p+1)}\,, ~~~~~   d \star   J^{(p)}_0 =0 
\end{eqnarray}
Making use of the PCO formalism we write the corresponding conserved charge as
\begin{eqnarray}
\label{cicC}
Q({\cal C}) =  \int_{{\cal M}^{(n-1)}} \star  J^{(p)}_0 \wedge \mathbb{Y}^{(p)}_{\cal C} = \int_{{\cal M}^{(n-1)}}  J^{(p)}_0 \wedge  \mathbb{Y}^{(n-1-p)}_{\cal C} 
\end{eqnarray}
where we have defined $ \mathbb{Y}^{(n-1-p)}_{\cal C} \equiv \star  \mathbb{Y}^{(p)}_{\cal C} $. The PCO $\mathbb{Y}^{(p)}_{\cal C}$ is a $p$-form which localizes the integral on a submanifold ${\cal C} \subset {\mathcal M}^{(n-1)}$ 
with dimension $(n-1-p)$ or equivalently spatial codimension $p$. This operator is closed but not exact respect to the space differential $d = \sum_{i=1}^n dx^i \partial_i$. Moreover, any variation inside the class of homological equivalent hypersurfaces in ${\mathcal M}^{(n-1)}$ is $d$-exact, as recalled in equation \eqref{PCOAB}. 

As a consequence of the last property the charge $Q$  is independent of the particular choice of ${\cal C}$. In fact, given two homologically equivalent hypersurfaces ${\cal C}$ and ${\cal C}'$ the corresponding PCOs differ by an exact term $  \mathbb{Y}^{(p)}_{{\cal C}'} = \mathbb{Y}^{(p)}_{\cal C} +d \Omega^{(p-1)}$. Therefore, we easily have 
\begin{eqnarray}
\label{cicE}
\hspace{-0.5cm} Q({\cal C}') - Q({\cal C}) =   \int_{{\cal M}^{(n-1)}} \star J^{(p)}_0 \wedge \left( \mathbb{Y}^{(p)}_{{\cal C}'} - \mathbb{Y}^{(p)}_{\cal C} \right) = \int_{{\cal M}^{(n-1)}} (d\star J^{(p)}_0) \wedge \Omega^{(p-1)}  =0
\end{eqnarray}
where we have integrated by parts the differential and used the second conservation law in (\ref{cicB}).

Using the first equation in \eqref{cicB} the charge conservation reads in general
\begin{eqnarray}
\label{cicD}
\partial_0 Q({\cal C}) =  \int_{{\cal M}^{(n-1)}} \!
\left(d  (\star  J^{(p+1)}) \wedge \mathbb{Y}^{(p)}_{\Sigma} + \star  J^{(p)}_0 \wedge \partial_0 \mathbb{Y}^{(p)}_{\Sigma} \right) \overset{?}{=} 0
\end{eqnarray}
While the first term is automatically vanishing due to the space-closure of the PCO, the vanishing of the second term deserves a separate discussion. In fact, it occurs not only when  $\partial_0 \mathbb{Y}^{(p)}_{\cal C}$ is zero but more generally when it is $d$-exact.
The first case corresponds to ordinary conserved charges defined on static hypersurfaces for which the defining equations do not depend on $x^0$. It is interesting to note that if  $\partial_0 \mathbb{Y}^{(p)}_{\cal C} = 0$ then the PCO is closed also respect to the spacetime differential $\hat d =  \sum_{i=0}^n dx^i \partial_i$.
In the more general case in which $\partial_0 \mathbb{Y}^{(p)}_{\cal C}$ is not vanishing but $d$-exact\footnote{The origin of this property is better understood if we embed $\mathbb{Y}^{(p)}_{\cal C}$  into a spacetime $p$-form $\tilde{ \mathbb{Y}}^{(p)}_{\cal C} =  \mathbb{Y}^{(p-1)}_0  dx^0 +  \mathbb{Y}^{(p)}_{\cal C}$. It is then easy to prove that requiring  $\hat{d} \, \tilde{\mathbb{Y}}^{(p)}_{\cal C} =0$ where $\hat{d}$ is the spacetime differential  implies $d  \mathbb{Y}^{(p)}_{\cal C} =0$ and $\partial_0 \mathbb{Y}^{(p)}_{\cal C} = - d \mathbb{Y}^{(p-1)}_0$.} the PCO depends non-trivially on $x^0$ and the corresponding  hypersurface becomes a dynamical object whose shape varies in time. However, the $Q$ charge is still conserved thanks to the second equation in \eqref{cicB}, as long as the hypersurface variations do not meet singularities.

As a clarifying example we consider the simple representative 
\begin{eqnarray} \label{example}
\mathbb{Y}^{(p)}_{\cal C} = \prod_{i=1}^{p} \delta(\phi_i) d \phi_i 
\end{eqnarray}
where $\phi_i (x^1, \dots , x^p) =0$, are the $p$ algebraic equations identifying the geometrical locus of the codimension-$p$ surface ${\cal C}$. Since for the time being we take the $\phi$'s to be independent of the time coordinate this defines a static PCO. It is easy to verify that $d \mathbb{Y}^{(p)}_{\cal C} =0$ but it is not exact. 

Now, evaluating $\mathbb{Y}^{(n-1-p)}_{\cal C} = \star \mathbb{Y}^{(p)}_{\cal C}$ and inserting it in \eqref{cicC} the corresponding conserved charge takes the form
\begin{eqnarray}
\label{cicI}
Q({\cal C}) =  \int_{{\cal M}^{(n-1)}}   J^{(p)}_0\wedge 
\prod_{i=1}^{p} \delta(\phi_i) \, \iota_{X_1} \dots \iota_{X_{p}} \, d^{(n-1)} \! x
\end{eqnarray}
where $X_1, \dots , X_p$ are vectors normal to the hypersurface $\Sigma$. Intuitively the contraction of the volume form along 
these vectors removes the dependence from $\prod_i d\phi_i$.
If we move the contractions on the $p$-form current, we use the product of Dirac delta functions to localize the integral  and integrate in the directions orthogonal to the hypersurface we finally obtain
\begin{eqnarray}
\label{cicF}
Q({\cal C}) &=& \int_{\cal C} \left(\iota_{X_1} \dots \iota_{X_{p}} J^{(p)}_0 \right) \, d^{(n-1-p)} \! x =
 \int_{\cal C}  J^{(p)}_{0, i_1 \dots i_{p}} X_1^{i_1} \dots   X_{p}^{i_{p}} \, d^{(n-1-p)} \! x
\end{eqnarray}
This coincides with the expression for the conserved charges that can be found in the literature \cite{Seiberg:2020wsg}. 

More generally, we consider a PCO of the form \eqref{example} but now corresponding to locus equations $\phi_i (x^0,x^1, \dots , x^p) =0$ which depend also on the time coordinate $x^0$. Precisely, we define
\begin{eqnarray}
\tilde{\mathbb{Y}}^{(p)}_{\cal C} = \prod_{i=1}^{p} \delta(\phi_i) \hat{d} \phi_i = \prod_{i=1}^{p} \delta(\phi_i) \partial_0 \phi_i dx^0 + \prod_{i=1}^{p} \delta(\phi_i) d \phi_i 
\equiv  {\mathbb{Y}}^{(p-1)}_{0} dx^0 + {\mathbb{Y}}^{(p)}_{\cal C}
\end{eqnarray} 
where ${\mathbb{Y}}^{(p)}_{\cal C}$ is the previous PCO \eqref{example} referred to a spatial slice at fixed $x^0$ \footnote{This definition assumes the possibility to foliate the spacetime manifold with space-like submanifolds and breaks diffeomorphism invariance in $n$ dimensions.}. 
It is easy to verify that this operator is $\hat{d}$-closed but not exact, and its $\hat{d}$-closure is equivalent to $d {\mathbb{Y}}^{(p)}_{\cal C} = 0 $ and $\partial_0 {\mathbb{Y}}^{(p)}_{\cal C} = - d  {\mathbb{Y}}^{(p-1)}_{0}$. Therefore, as discussed above, charge \eqref{cicC} when defined in terms of ${\mathbb{Y}}^{(p)}_{\cal C}$ is conserved.

\subsection{$(p+1)$-form Supercurrents}\label{supercurrents}

The geometric formulation of conservation laws discussed above allows for a straightforward generalization to supermanifolds. Here we discuss the construction of conserved tensorial supercurrents  in a supermanifold $\mathcal{SM}^{(n|m)}$. 

We begin by considering a $(p+1)$-tensorial abelian supercurrent described by the superform 
\begin{eqnarray}
\label{scurA}
\hat J^{(p+1|0)} = \S_{k=0}^{p+1} \; J_{a_1 \dots a_k \a_{k+1} \dots \a_{p+1}} \, V^{a_1} \dots V^{a_k}  \, \psi^{\alpha_{k+1}} \dots \psi^{\alpha_{p+1}}
\end{eqnarray}
where $V^a = dx^a + \theta \g^a d \theta$ and $\psi^\a = d \theta^\a$ are the supervielbeins. We recall that components $J_{a_1 \dots a_k \a_{k+1} \dots \a_{p+1}}$ are functions of the $(x,\theta)$ coordinates, thus they are superfields. 

The conservation law is expressed as usual as $d^{\hat \dagger} \hat J^{(p+1|0)} =0$, with the conjugate differential given by $d^{\hat \dagger} = \hat\star d \hat\star$, being $\hat\star$ the Hodge dual on the entire supermanifold defined in appendix (\ref{hodge})\footnote{The complete theory is developed in \cite{Castellani:2014goa,Castellani:2015ata,Catenacci:2016qzd}.}.

Ordinary vector currents in superspace are obtained by setting $p=0$. The corresponding conservation law reads
\begin{eqnarray}
\label{scurB}
0 = \hat\star d (\hat\star \hat J^{(1|0)}) &=& \hat\star d \left( J_a g^{a b_1} \epsilon_{b_1 \dots b_n} V^{b_2} \dots V^{b_n} \delta^m(\psi) + 
J_{\alpha} g^{\alpha \beta } V^1 \dots V^n \iota_{\beta} \delta^m(\psi) \right) \nonumber \\
&=& \hat\star \left( \partial_c J_a V^c g^{a b_1} \epsilon_{b_1 \dots b_n} V^{b_2} \dots V^{b_n} \delta^m(\psi) + 
 D_\gamma J_{\alpha} g^{\alpha \beta } \psi^\gamma  V^1 \dots V^n \iota_{\beta} \delta^m(\psi) \right) \nonumber \\
 &=& 
  \hat\star \left( \partial_a J^a  -
 D_\beta J_{\alpha} g^{\alpha \beta }  \right) V^{1} \dots V^{n} \delta^m(\psi)  \nonumber \\
 &=& \left( \partial_a J^a  +
 D_\a J^{\alpha} \right) = D_\a \Big(J^\alpha + \gamma_a^{\a\b} D_\b J^a \Big) 
\end{eqnarray}
where in the last line we have used the superspace identity $\partial_a = \gamma_a^{\a\b} D_\a D_\b$. 
The quantity $\tilde{J}^\a = (J^\alpha + \gamma_a^{\a\b} D_\b J^a )$ is the most general expression for a $U(1)$ supercurrent in superspace and $D_\a \tilde{J}^\a = 0$ is the standard conservation law. 

We now study supercurrents \eqref{scurA} for $p>0$.  For simplicity we consider the $p=1$ case and compute the action of $d^{\hat \dagger}$ on  
\begin{eqnarray}
\hat J^{(2|0)} =  J_{ab} V^a V^b + J_{a\b} V^a \psi^\beta + J_{\a\b} \psi^\a \psi^\b
\end{eqnarray}
The result is a $1$-superform which can be explicitly obtained by the following chain of identities
\begin{eqnarray}
\label{scurB}
 && \hspace{-0.5cm} d^{\hat \dagger} \hat J^{(2|0)} = \hat\star d (\hat\star \hat J^{(2|0)}) =\nonumber \\
&=& \hat\star \, d \Big( J_{a b} \, g^{a c_1} g^{b c_2} \epsilon_{c_1 c_2 c_3 \dots c_n} V^{c_3} \! \dots \! V^{c_n} \delta^m(\psi) \nonumber \\
&&~~~+ 
J_{a \b} g^{a c_1} g^{\b \gamma} \epsilon_{c_1 c_2 \dots c_n} V^{c_2} \! \dots \! V^{c_n} \iota_\gamma  \delta^m(\psi) 
+J_{\alpha \beta} g^{\alpha \gamma_1 } g^{\beta \gamma_2 } V^1 \! \dots \!  V^n \iota_{\gamma_1} \iota_{\gamma_2} \delta^m(\psi) \Big)  \nonumber \\
&=& \hat\star 
\Big( 
\partial_c J_{ab}   g^{a c_1}  g^{b c_2} \epsilon_{c_1 c_2 c_3\dots c_n} V^c V^{c_3} \! \dots \!  V^{c_n} \delta^m(\psi) +\partial_c J_{a\beta}  g^{a c_1} g^{\b \gamma} \epsilon_{c_1 c_2 \dots c_n} V^c V^{c_2} \! \dots \!  V^{c_n} \iota_\gamma  \delta^m(\psi)  \nonumber \\
&&~~~ + D_\delta J_{a \beta} g^{a c_1} g^{\beta \gamma } \psi^\delta \epsilon_{c_1 c_2 \dots c_n} V^{c_2} \! \dots \!  V^{c_n} \iota_\gamma \delta^m(\psi) 
+ D_\gamma J_{\a \beta} \psi^\gamma g^{\a \gamma_1} g^{\alpha \gamma_2} 
 V^1 \! \dots \!  V^n \iota_{\gamma_1}\iota_{\gamma_2} \delta^m(\psi) \Big)  \nonumber \\ 
 &=& \hat\star 
\Big( 
\partial_c J_{ab}   g^{a c_1}  g^{b c_2} \epsilon_{c_1 c_2 c_3\dots c_n} V^c V^{c_3} \! \dots \!  V^{c_n} \delta^m(\psi)  + 
\partial_c J_{a\beta}  g^{a c_1} g^{\b \gamma} \epsilon_{c_1 c_2 \dots c_n} V^c V^{c_2} \! \dots \! V^{c_n} \iota_\gamma  \delta^m(\psi) 
 \nonumber \\
&&~~~ - D_\gamma J_{a \beta} g^{a c_1} g^{\beta \gamma } \epsilon_{c_1 c_2 \dots c_n} V^{c_2}\! \dots \! V^{c_n} \delta^m(\psi) 
 - D_\gamma J_{\a \beta}  g^{\a \gamma} g^{\beta \gamma_2} 
 V^1\! \dots \!  V^n \iota_{\gamma_2} \delta^m(\psi) \Big) \nonumber \\ 
 &=& 
 (\partial^a J_{ab} - D^\a J_{b \alpha}) V^b +  (\partial^a J_{a \beta} - D^\a J_{\alpha \beta}) \psi^\b \nonumber\\
 &=& 
 D^\a(\gamma^a_{\a\b} D^\b J_{ab} -  J_{b \alpha}) V^b +  D^\a (\gamma^a_{\a\gamma}  D^\gamma J_{a \beta} - J_{\alpha \beta}) \psi^\b
 \end{eqnarray}
In the last line we have used the superspace identity $\partial_a = \gamma_a^{\a\b} D_\a D_\b$. 
Now, if we define the two currents $\tilde{J}_{\a b} = (J_{\alpha b} - (\gamma^a)_\a^{\b} D_\b J_{a b} )$ and 
$\tilde{J}_{\a \b} = (J_{\alpha \b} - (\gamma^a)_{(\a}^{\gamma} D_\gamma J_{|a| \beta)})$, the conservation law for a $(2|0)$ supercurrent, 
$d^{\hat \dagger} J^{(2|0)} =0$, turns out to be equivalent to the two conservation laws
\begin{eqnarray}
\label{miiA}
D^\a \tilde{J}_{\a b} =0\,, ~~~~~~
D^\a \tilde{J}_{\a \b} =0
\end{eqnarray}

This result is easily generalizable to $(p+1)$-supercurrents with $p>2$. Since $d^{\hat \dagger}$ maps $(p+1)$-superforms into $p$-superforms, the condition
$d^{\hat \dagger} \hat J^{(p+1|0)} =0$ gives rise to $(p+1)$ conserved supercurrents. Following the same procedure highlighted above one can find the explicit expressions of the $(p+1)$ currents in terms of the $\hat J^{(p+1|0)}$ components. 

As done in ordinary manifolds, we can split the supercurrent in its time and spatial components $\hat J^{(p+1|0)} = (J_0^{(p|0)} , J^{(p+1|0)})$, so that its conservation law reads 
\begin{eqnarray}
\label{scurD}
\partial_0 J^{(p|0)}_0 = d^\dagger J^{(p+1|0)}\,, ~~~~ d^\dagger  J_0^{(p|0)}=0
\end{eqnarray}
where now $d^\dagger = \star d \star$, being $\star$ the Hodge dual in the constant time slice $\mathcal{SM}^{(n-1|m)}$. In particular, it satisfies identities \eqref{app:star} with $n \to n-1$.

In order to define a conserved supercharge associated to this supercurrent we use the PCO technique and write \begin{eqnarray}
\label{scurE}
Q({\mathbb C}) = \int_{\mathcal{SM}^{(n-1|m)}} \star J_0^{(p|0)}\wedge \mathbb{Y}^{(p|0)}_{\mathbb C} = \int_{{\mathcal{SM}}^{(n-1|m)}}   J_0^{(p|0)}\wedge  \mathbb{Y}^{(n-1-p \, |m)}_{\mathbb C}
\end{eqnarray}
where we have defined $\mathbb{Y}^{(n-1-p \, |m)}_{\mathbb C}= \star \mathbb{Y}^{(p|0)}_{\mathbb C}$. Here $\mathbb{Y}^{(p|0)}_{\mathbb C}$ is the PCO localizing the integral on a  spatial submanifold ${\mathbb C}$ of dimensions $(n-1-p \,|m)$.  
Once the integration on the supermanifold is performed the charge does not depend upon the fermionic coordinates. 

The $Q$ charge satisfies the conservation law $\pa_0 Q=0$ as a consequence of identities \eqref{scurD}, which in turn encode $(p+1)$ conservation laws. Moreover, with a reasoning similar to the one used in the bosonic case (see eq. \eqref{cicE}), it is easy to prove that $Q({\mathbb C})$ does not depend on the particular choice of the surface, thanks to the second constraint in \eqref{scurD}.

\subsection{Charged Defects}

We now investigate which are the physical objects that are charged under $p$-form symmetries generated by $(p+1)$-form conserved (super)currents. We start discussing the bosonic case, basically reviewing in our language results of \cite{Gaiotto:2014kfa,Seiberg:2019vrp}, and then generalize to the tensorial supercurrents that we have just constructed.

Objects that are charged under $Q({\cal C})$ defined in \eqref{cicC} are Wilson-type operators of the form \eqref{AKAFgen} with $\G =  i \int_{\cal M} B^{(p)} \wedge {\mathbb{Y}}_\s^{(n-p)}$, being 
$\s$ a dimension-$p$ hypersurface in ${\cal M}$ \cite{Gaiotto:2014kfa,Seiberg:2019vrp}. In particular, for $p=2$ the charged objects are the Wilson surfaces that we have discussed in section \ref{sect:bosonicWS}. For generic $p$, given the $Q$ charge in \eqref{cicC} we can write
\begin{eqnarray}\label{cicG}
e^{i\beta Q({\cal C})} \, W_p[\s] \, e^{-i\beta Q({\cal C})} = e^{i \beta  I(\s, {\cal C})} \, W_p[\s] 
\end{eqnarray}
where $I(\s, {\cal C})$ is the linking number of $\s$ and ${\cal C}$, or equivalently the intersection number of $\s$ and a submanifold ${\cal B}$ whose boundary is ${\cal C}$. It is a topological invariant that counts the number of points in ${\cal M}^{(n)}$ at which $\s$ intersects ${\cal B}$. In our formalism this quantity can be expressed in a simple manner in terms of the corresponding PCOs (for a general discussion see appendix \ref{app:linking}). We first express the $Q$ charge as an integral of a top form on the entire manifold ${\cal M}^{(n)}$ by including the PCO in \eqref{TCDt}. Exploiting the PCOs closure we can write\footnote{As already mentioned, in writing $\mathbb{Y}^{(p)}_{\cal C} \wedge \mathbb{Y}^{(1)} = \hat d \Omega_{\cal C}^{(p)}$ there is no contradiction with the general statement that PCOs are not exact, since $\Omega_{\cal C}^{(p)}$ contains a distribution with non-compact support.}  
\begin{eqnarray}
Q({\cal C}) = \int_{{\cal M}^{(n)}} \star J^{(p)}_0 \wedge \mathbb{Y}^{(p)}_{\cal C} \wedge \mathbb{Y}^{(1)} = \int_{{\cal M}^{(n)}} \star  J^{(p)}_0 \wedge \hat d \left( \mathbb{Y}^{(p)}_{\cal C} \Theta(x^0) \right) \equiv \int_{{\cal M}^{(n)}} \star J^{(p)}_0 \wedge \hat d \Omega_{\cal C}^{(p)} \nonumber \\
\end{eqnarray}
Using the general result in appendix \ref{app:linking}, it follows that the linking number in eq. \eqref{cicG} is explicitly given by\footnote{In general, this is  known as {\it cup product}, see for example \cite{botttu}.} 
\begin{eqnarray}\label{cicH}
I(\s, {\cal C}) = \int_{{\cal M}^{(n)}} \Omega_{\cal C}^{(p)} \wedge {\mathbb{Y}}_\s^{(n-p)}
\end{eqnarray}
It is consistently defined as the integral on the whole manifold of a top form as a consequence of the fact that the dimensions of the hypersurface operator is the same as the tensorial degree of the symmetry. 

We now move to supermanifolds and argue  that objects charged under  $Q({\mathbb C})$ in \eqref{scurE} are $(p|0)$-dimensional hypersurface operators $W_p[\S]$ defined in \eqref{AKAFgen}. 
Generalising to supermanifolds  the construction in eqs. (\ref{cicG}, \ref{cicH}), we obtain that the action of the $Q$-charge corresponding to a $(p|0)$-form symmetry on an hypersurface operator of dimension $(q|0)$ reads
\begin{eqnarray}\label{cicJ}
e^{i\b Q({\mathbb C})} \, W_q[\S] \, e^{-i\b Q({\mathbb C})} = e^{i\b I(\S, {\mathbb C})} W_q[\S]
\end{eqnarray}
where $I(\S, {\mathbb C})$ is the {\em super-linking number} between the supermanifolds $\S$ and ${\mathbb C}$, defined in appendix \ref{app:linking}. In order to express it in terms of the PCOs, we first extend the integral defining $Q({\mathbb C})$ to the whole supermanifold by using the time PCO in \eqref{TCDt}, now generalized to a $(1|0)$-form in supermanifold
\begin{eqnarray}
Q({\mathbb C}) &=& \int_{{\cal{SM}}^{(n|m)}} \! \! \!  \star  J^{(p|0)}_0 \wedge \mathbb{Y}^{(p|0)}_{\mathbb C} \wedge \mathbb{Y}^{(1|0)} = \int_{{\cal{SM}}^{(n|m)}} \! \! \! \star J^{(p|0)}_0 \wedge \hat d \left( \mathbb{Y}^{(p|0)}_{\mathbb C} \Theta(x^0) \right) \nonumber \\
&\equiv& \int_{{\cal{SM}}^{(n|m)}} \! \!   \star  J^{(p|0)}_0 \wedge \hat d \, \Omega_{\mathbb C}^{(p|0)}  
\end{eqnarray}
According to eq. \eqref{Bottu} it  then follows that the super-linking number appearing in \eqref{cicJ} is given by
\begin{eqnarray}\label{cicH2}
I(\S, {\mathbb C}) = \int_{{\cal{SM}}^{(n|m)}} \Omega_{\mathbb C}^{(p|0)} \wedge {\mathbb{Y}}_\S^{(n-q|m)}
\end{eqnarray}
where ${\mathbb{Y}}_\S^{(n-q|m)}$ has been introduced in \eqref{AKAFgen}. It is now easy to observe that this expression is non-vanishing only when it corresponds to the integral of a top form in supermanifold, that is only when $q=p$. Therefore, we conclude that objects charged under symmetries generated by $(p+1|0)$-form supercurrents  are Wilson-like $(p|0)$-dimensional defects in superspace.  
Choosing in particular $p=1$, we see that the WS operators that we have defined and studied in this paper describe physical defects that are charged under an abelian  $(1|0)$-form symmetry.

\sect{Conclusions and Perspectives}

We have generalized the geometric construction of (super)Wilson loops \cite{Cremonini:2020mrk} to the case of hypersurface operators. In particular, we have considered 2-form Wilson-like operators defined on $(2|0)$-dimensional supersurfaces described by a given embedding of bosonic and grassmanian coordinates in a supermanifold. 

In the case of Wilson Surfaces generated by the tensor multiplet of the six-dimensional $N=(2,0)$ SCFT and their generalization to include couplings to scalars, we have studied supersymmetry preserving constraints on the supersurface. By suitably choosing the cohomology representative in 
the set of Picture Changing Operators we have attempted a first classification of surfaces preserving different sets of supercharges.  Although we have worked in six dimensions, most of the results can be easily proved to be valid in other dimensions. 

In six dimensions we have also studied the behavior of WS under kappa-symmetry. We have found that kappa-symmetry invariance leads to the same constraints as supersymmetry invariance. Remarkably, the constraints for the invariance of the surface operator have a M-theory dual interpretation. They coincide with the constraints ensuring kappa-symmetry invariance of a static supermembrane. This observation hints to quest for a deeper geometrical understanding using M2/M5 systems, which might help in attempting a general classification of BPS (super)surfaces. In particular, we have found that the kappa-symmetry constraint in eleven dimensions, once dimensionally reduced, gives rise to the PBS conditions for the generalized WS in six dimensions, in analogy with what happens for Wilson-Maldacena loops. 

Since super-hypersurface operators  describe objects that should be charged under global symmetries generated by tensorial conserved supercurrents, in the last part of the paper we have studied tensorial conservation laws in superspace. To this end, we have first reformulated the known bosonic $p$-form conservation laws in  geometric language, in terms of forms and PCOs. Then we have generalized this construction to the supersymmetric case, simply by replacing forms defined in manifolds with superforms living in supermanifolds. In particular, the geometric formulation of conservation laws for p-form supercurrents has required the use of a Hodge dual suitably extended to supermanifolds. The main result is that the super-conservation law for a $(p|0)$-tensor supercurrent leads to $p$ independent conservation laws. The physical meaning of these multiple conservation laws has still to be deeply investigated. 
We have finally discussed the relation between super-hypersurface operators and tensorial (super)symmetries. In particular, the assignment of a $p$-form charge to a $p$-dimensional hypersurface operator has required the generalization to supermanifolds of the concept of linking number. 

Our construction can be generalized to define  $(p|m)$-integral currents, that is conserved  integral forms, or more generally $(p|q)$-form currents with $0 < q < m$ described by conserved pseudo-forms. The physical meaning of this conservation laws and the corresponding exotic symmetries has still to be understood and will be discussed elsewhere \cite{new}. 

Furthermore, our approach can be exploited to generalize to supermanifolds the recent formulation of a continuum field theory for probe particles and dipoles with reduced mobility (fractons and lineons) \cite{Pretko:2016kxt, Pretko:2018jbi, Gromov:2018nbv, Seiberg:2020bhn,Seiberg:2020wsg,Seiberg:2020cxy}. In the bosonic case, if a dipole symmetry is gauged by introducing a corresponding tensor gauge field, surface operators can be defined which probe the motion of charged dipole particles in such a background. Since gauge invariance highly constrains their motion, these quantities can be used to describe particles with reduced mobility (lineons) \cite{Seiberg:2020bhn,Seiberg:2020wsg,Seiberg:2020cxy}. The introduction of dipole (or more generally multipole) supercurrents in superspace leads immediately to the possibility of generalizing this physical construction to supersymmetric theories. Gauging a tensorial symmetry generated by a conserved multipole supercurrent
leads to the introduction of a tensorial gauge superfield, and the corresponding Wilson-like extended objects should naturally describe new states of matter with reduced motion in such a super-background ({\em superfractons} and {\em superlineons}). The physical properties of such objects and the role of supersymmetry in this game are presently under investigation \cite{new}.

As the last remark, we recall that we have considered only abelian operators, that is WS or higher dimensional operators constructed with abelian tensor forms. Accordingly, we have focused only on abelian tensorial supercurrents. It would be interesting to generalize our construction to the non-abelian case. As already mentioned, the main problem is to find a consistent definition of normal-ordered exponential when the manifold on which the Wilson-type operator is localized has dimension greater than one. Some recent proposals can be found in \cite{Chepelev:2001mg,Hofman:2002ey,Baez:2010ya,Ho:2012nt,Kim:2019owc}. We plan to go back to this problem in a near future.

\vskip 15pt
\section*{Acknowledgements}

This work has been partially supported by Universit\`a del Piemonte Orientale research funds, by Italian Ministero dell'Universit\`a e della Ricerca (MIUR), and by Istituto Nazionale di Fisica Nucleare (INFN) through the ``FieLds And Gravity'' (FLAG) and ``Gauge theories, Strings, Supergravity'' (GSS) research projects.

\newpage
\appendix
\section{Conventions in six dimensions} \setcounter{equation}{0} \label{appendix}

In this appendix we collect some notations and formulae for the six-dimensional $N=(2,0)$ superspace. 
We refer for example to \cite{Howe:1983fr,Bergshoeff:1985mz,Claus:1997cq,Ferrara:2000xg} for a complete description. 

\noindent
We begin by fixing some index notation. We work in Minkowski spacetime with mostly plus signature. 

\noindent
$\ast$  We use $\a, \b, \gamma =1, \dots 4$ to denote 
$SU^*(4)$ (the spinorial representation of $SO(1,5)$ Lorentz group) indices. Upper/lower indices correspond to right-handed/left-handed spinorial indices, respectively. 

\noindent
$\ast$ We use the middle Greek letters $\mu, \nu, \dots =0, \dots, 5$ to denote vector indices. 

\noindent
$\ast$  Finally, we use capital latin letters $A,B,C, \dots,=1, \dots, 4$ to denote 
$USp(4)$ indices of R-symmetry group. The antisymmetric matrix $\Omega_{AB}$ is the symplectic form 
preserved by $USp(4) \sim SO(5)$ group which can be put (using Darboux coordinates) in the form 
\begin{eqnarray}
\label{SS6B}
\Omega_{AB} = \left(
\begin{array}{cc}
 0 &  {\mathcal I}   \\
 -{\mathcal I}  &   0
\end{array}
\right)
\end{eqnarray}
where ${\mathcal I}$ is the $2\times2$ identity matrix. The upper-index matrix $\Omega^{AB}$ is defined by the condition $\Omega_{AB} \Omega^{BC} = - \delta_A^{~C}$ and is formally equal to $\Omega_{AB}$. The matrix  $\Omega$ is used 
 to raise and lower R-symmetry  indices as $\lambda^A =  \Omega^{AB} \lambda_B\,, \lambda_A =  \lambda^B \Omega_{BA}$. 
 

Explicity, we use the following Dirac Matrix representation
 $\Gamma_M =  \{\Gamma_\mu,\Gamma_r\}$ (with $\mu=0,\dots,5$ and $r=1,\dots,5$ and the chirality 
 matrix $\Gamma_7 = \Gamma_0 \Gamma_1 \Gamma_2 \Gamma_3 \Gamma_4 \Gamma_5$
given (in the chiral basis)
\begin{eqnarray}
\label{MAA}
\Gamma_\mu =
\left(
\begin{array}{cc}
 0 & \bar\gamma_\mu  \\
\gamma_\mu  & 0 
\end{array}
\right) \otimes I_4
\,, ~~~~~~~
\Gamma_r =
\left(
\begin{array}{cc}
 -I_4 & 0  \\
0  &  I_4
\end{array}
\right) \otimes \hat{\gamma}_r
\,, ~~~~~~~
\Gamma_7 =
\left(
\begin{array}{cc}
 -I_4 & 0  \\
0  &  I_4
\end{array}
\right) \otimes I_4
\end{eqnarray}
where 
\begin{eqnarray}
\label{MAB}
\bar\gamma_\mu \gamma_\nu + \bar\gamma_\nu \gamma_\mu = 2 \eta_{\mu \nu}\,, ~~
\gamma_\mu \bar\gamma_\nu + \gamma_\nu \bar\gamma_\mu = 2 \eta_{\mu \nu}\,, ~~
\{\hat\gamma_r, \hat\gamma_s\} = 2 \delta_{rs}\,, ~~~[\gamma_\mu, \hat\gamma_r]=0
\end{eqnarray}
Explicitly, we have
\begin{eqnarray}
\label{gm}
\begin{array}{lll}
 \hspace{-0.3cm} \gamma_0 = \bar\gamma_0= i I_2 \otimes I_2\,,  
 & 
 \hspace{-0.3cm} \gamma_1 = - \bar\gamma_1= - i \sigma_1 \otimes I_2\,,  
 &   
 \hspace{-0.3cm} \gamma_2= - \bar\gamma_2 = - i \sigma_2 \otimes I_2 \,,  
 \\
\hspace{-0.3cm}  \gamma_3 = -\bar\gamma_3= i \sigma_3 \otimes \sigma_1\,,  
 &    
 \hspace{-0.3cm} \gamma_4 = -\bar\gamma_4= i \sigma_3 \otimes \sigma_2\,, 
 &   
 \hspace{-0.3cm} \gamma_5 = -\bar\gamma_5= - i \sigma_3 \otimes \sigma_3\,,  
 \\
\hspace{-0.3cm}  \hat\gamma_1 = \sigma_1 \otimes \sigma_2\,, ~~~ \hat\gamma_2 = \sigma_2 \otimes \sigma_2\,, 
 & 
\quad  \hat\gamma_3 = \sigma_3 \otimes \sigma_2\,, ~~~ \hat\gamma_4 = I_2 \otimes \sigma_1\,, 
 &   
 \quad \hat\gamma_5 = I_2 \otimes \sigma_3\,, 
\\
\hspace{-0.3cm} c = - c^T= \sigma_1 \otimes i \sigma_2\,, 
&
\Omega = i \sigma_2 \otimes I_2 
\end{array}
\end{eqnarray}
where $c$ is the charge conjugation matrix and $\Omega$ is the antisymmetric tensor $\Omega_{AB}$

\vskip 5pt
The six-dimensional $N=(2,0)$ superspace ${\mathcal SM}^{(6|16)}$ is described by the following coordinates
\begin{eqnarray}
\label{SS6C}
x^{[\a\b]} = \gamma^{[\a\b]}_\mu x^\mu\,, ~~~~ \theta^\a_A  
\end{eqnarray}
subject to the Majorana-Weyl pseudoreality condition $\overline \theta^A_\b = \Omega^{AB} \theta^\a_B c_{\a\b}$, with $c_{\a\b}$ the charge conjugation matrix. 
The invariant 1-forms are then given by  
\begin{eqnarray}
\label{SS6D}
V^{[\a\b]} = d x^{[\a\b]} + \theta^{[\a}_A \Omega^{AB} d\theta^{\b]}_B\,,  ~~~~~
\psi^\a_A = d\theta^\a_A
\end{eqnarray}
(notice that $\theta^{[\a}_A \Omega^{AB} \theta^{\b]}_B =0$), while the basic  Maurer-Cartan are 
\begin{eqnarray}
\label{SS6E}
d V^{[\a\b]} = \psi^{[\a}_A  \Omega^{AB} \psi^{\b]}_B\,, ~~~~~ d  \psi^{\a}_A =0 
\end{eqnarray}
The superderivatives are defined as 
\begin{eqnarray}
\label{SS6F}
D^A_\a = \frac{\partial}{\partial \theta^\a_A} + i\Omega^{AB} \theta^\b_B \partial_{\a\b}\,, ~~~~~~
\{D^A_\a , D^B_\b \} = 2 i \Omega^{AB} \gamma_{\a\b}^\mu \partial_\mu 
\end{eqnarray}
Similar definitions hold for the $Q^A_\a$ supercharges, 
\begin{eqnarray}
\label{SS6H}
Q^A_\a = \frac{\partial}{\partial \theta^\a_A} - i\Omega^{AB} \theta^\b_B \partial_{\a\b}\,, ~~~~~~
\{Q^A_\a , Q^B_\b \} = 2 i \Omega^{AB} \gamma_{\a\b}^\mu \partial_\mu 
\end{eqnarray}
The generators $L^{AB}$ of $USp(4)$ form the algebra
\begin{eqnarray}
\label{SS6G}
[L^{AB}, L^{CD}] = \Omega^{A (C} L^{D) B} +  \Omega^{B (C} L^{D) A}\,, ~~~~~ [L^{AB}, D^C_\a] = - \Omega^{C (A} D^{B)}_\a   
\end{eqnarray}

{\em Kappa symmetry} is a superdiffeomorphism generated by the spinorial field $\tilde{\kappa} = \kappa^\alpha_A D^A_\alpha$. It acts on the coordinates as
\begin{eqnarray} \label{Ktransfs}
	\delta_{\widetilde \k} \theta^\alpha_A &=& {\cal L}_{\widetilde \k} \theta^\a_A =  \k^\beta_B \iota_\beta^B \psi^\alpha_A = \k^\alpha_A  \nonumber \\
	\delta_{\widetilde \k} x^{[\alpha \beta]} &=& {\cal L}_{\widetilde \k} x^{[\alpha \beta]} = \k^\gamma_C \iota_\gamma^C V^{[\alpha \beta]} + \k^\gamma_C \iota_\gamma^C i \Omega^{AB} \theta_B^\beta \psi^\alpha_A = i \k^\alpha_A \Omega^{AB} \theta_B^\beta
\end{eqnarray}
where we have used the contractions $\iota_{\tilde \kappa}$ of the six dimensional supervielbeins as given by
\begin{eqnarray}\label{KSB} 
\iota_{\tilde{\kappa}} \psi^{\alpha}_A &=& \kappa^\beta_B \iota^B_\beta \psi^\alpha_A = \kappa^\beta_B \delta^\alpha_\beta \delta^B_A = \kappa^\alpha_A \ , \\
	\nonumber \iota_{\tilde{\kappa}}  V^{[\alpha \beta]} &=& \kappa^\gamma_C \left[ -i \Omega^{CD} \theta_D^\delta \delta^{[\alpha}_\gamma \delta^{\beta]}_\delta + i \theta^{[\alpha}_A \Omega^{AB} \delta^C_B \delta^{\beta]}_\gamma \right] = -i \kappa^{[\alpha}_A \Omega^{AB} \theta^{\beta]}_B + i \kappa_B^{[\beta} \theta_A^{\alpha]} \Omega^{AB} = 0
\end{eqnarray}

\section{Hodge operator in Supermanifolds}\label{hodge} \setcounter{equation}{0}

In this appendix we briefly recall how to define the Hodge operator on a supermanifold. The general construction and a few applications can be found in \cite{Castellani:2015ata}. 

The easiest way to generalize the Hodge operator to a supermanifold is to start from the representation of the usual Hodge operator in manifolds in terms of an \emph{odd Fourier transform} of the \virgolette differential part" of a given form. Precisely, given a $p$-form $\displaystyle \omega^{(p)} (x,dx)$ in a $n$-dimensional manifold, we introduce $n$ Grassmann variables $\eta^{a=1, \dots , n}$. It is then easy to see that the Hodge dual $(n-p)$-form $\star \omega^{(p)}(x,dx)$ can be defined as Berezin-integral over the $\eta$'s 
\begin{equation}
	\star \omega^{(p)} (x,dx)= i^{p^2-n^2} \mathcal{F} [\omega] = i^{p^2-n^2} \int_{\eta} \omega^{(p)} \left( x, \eta \right) e^{i dx \cdot \eta}
\end{equation}
where the integrand is the original $p$-form with the $dx$'s substituted by the $\eta$'s, the \virgolette Fourier kernel" is given in terms of its (finite) series expansion and $dx \cdot \eta \equiv dx^a g_{ab} \eta^b$, being $g_{ab}$ the metric tensor of the manifold. The overall coefficient is chosen in order to reproduce the usual identity $\star \star \omega^{(p)} = (-1)^{p(n-p)} \omega^{(p)}$.

This construction can be easily generalizable to define  the Hodge dual on a  $(n|m)$-dimensional supermanifold $\mathcal{SM}$. Introducing odd variables 
$\eta^{a=1,\dots , n}$ and even variables $b^{\a =1, \dots, m}$, for a given a superform $\displaystyle \omega^{(p|q)} (x,dx, \theta , d \theta) \in \Omega^{(p|q)} \left( \mathcal{SM} \right)$ we define
\begin{equation}
	\star \omega = c \, \mathcal{F} [\omega] = c \int_{\eta,b} \omega \left( x, \eta , \theta , b \right) e^{i dx \cdot \eta + i d \theta \cdot b}
\end{equation}
where $c$ is a suitable normalisation coefficient. The scalar products in the exponential are given by $dx \cdot \eta = dx^a g_{ab} \eta^b$ and 
$d \theta \cdot b = d\theta^\a g_{\a \b} b^\b$, in terms of the metric tensor and an anti-symmetric tensor $g_{\a\b}$ (see \cite{Castellani:2015ata} for details). In particular, we observe that the Hodge operator sends superforms into integral forms and viceversa
\begin{equation}\label{app:star}
	\star : \Omega^{(p | 0)} \to \Omega^{(n-p | m)} \ , \qquad  \star : \Omega^{(p | m)} \to \Omega^{(n-p | 0)}
\end{equation}

As an easy example, let us consider the flat $\mathbb{R}^{(2|2)}$ manifold. In this case the products in the Fourier kernel are 
$dx \cdot \eta = dx^a \delta_{ab} \eta^b \equiv dx^a \eta_a$ and $d \theta \cdot b = d \theta^\alpha \epsilon_{\alpha \beta} b^\beta \equiv d\theta^\a b_\a$ and we have for example that
\begin{eqnarray}
 \hspace{-1cm} \mathcal{F} [dx^1] = \int_{b,\eta} \eta^1 e^{i dx^b \eta_b + i d \theta^\alpha b_\alpha} = \int_{b,\eta} \eta^1 (1 + i \, dx^2 \eta^2 ) e^{ i d \theta^\alpha b_\alpha} = i \, dx^2 \delta^{(2)} \left( d \theta \right) 
\end{eqnarray}
and
\begin{eqnarray}
\hspace{-0.5cm}  \mathcal{F} [\delta^{(2)} \left( d \theta \right)] = \int_{b,\eta} \delta^{(2)} \left( b \right) e^{i dx^b \eta_b + i d \theta^\alpha b_\alpha} = \int_{b,\eta} (1 + i \, dx^1 \eta^1) (1 + i \, dx^2 \eta^2) =  dx^1 dx^2
\end{eqnarray}

\section{Charge conservation in the extended manifold} \setcounter{equation}{0}

In this appendix we show that it is possible to rephrase the formalism presented in section 7 in terms of PCOs that contain an explicit dependence on the hypersurface parametrization. 

Given a $(n-1-p)$-dimensional hypersurface $\S$ embedded in the spatial manifold   $\mathcal{M}^{(n-1)}$, we parametrize it as $(\tau_1, \dots , \tau_p) \to x^a(\tau_1, \dots , \tau_p)$, $a=1, \dots, n-1-p$, and $\tau_i \in \Delta \subseteq \mathbb{R}^{n-1-p}$. We enlarge the manifold to $\mathcal{M}^{(n-1)} \times \Delta$, with coordinates $(x^a, \tau_i)$. In this framework, the charge $Q$ in \eqref{cicC} can be rewritten as
\begin{equation}\label{TCFE}
Q =  \int_{{\cal M}^{(n-1)}}  J_0 \, \wedge \mathbb{Y}_\S^{(n-1)} = \int_{{\cal M}^{(n-1)} \times \Delta}  
	J_0 \; d^{(n-1-p)} x \, \wedge \tilde{\mathbb{Y}}_\S^{(n-1)}
\end{equation}
where 
\begin{eqnarray}
 \tilde{\mathbb{Y}}_\S^{(n-1)} = \prod_{a=1}^{{n-1}} \delta \left( x^a - x^a \left( \tau_1 , \ldots , \tau_{n-1-p} \right) \right) \bigwedge_{a=1}^{n-1} \left( dx^a - \partial_i x^a d \tau^i \right) 
\end{eqnarray}
The operator $\tilde{\mathbb{Y}}^{(2n-2-p)} \equiv  d^{(n-1-p)} x\wedge \tilde{\mathbb{Y}}^{(n-1)}$ is the dual to the embedding $\tau_i \to (x^a(\tau_i) , \tau_i)$. 

According to the procedure described in section \ref{currents}, the charge conservation requires evaluating the Hodge dual of the PCO. Since now we work in the enlarged manifold, we define an enlarged Hodge dual $\star_{\tilde{g}}$ with respect to the metric $\tilde{g} = g \otimes \mathbb{I}$ on $\mathcal{M}^{(n-1)} \times \Delta$. It is then straightforward to evaluate
\begin{equation}\label{TCFF}
	\star_{\tilde{g}} {\tilde{\mathbb{Y}}^{(2n-2-p)}} = \prod_{a=1}^{{n-1}} \delta \left( x^a - x^a \left( \tau_1 , \ldots , \tau_{n-p-1} \right) \right) = \prod_{a=1}^{{n-1}} \frac{1}{n-1} \left( \iota_{\partial_{x^a}} - \left( \frac{\partial x^a}{\partial \tau^i} \right)^{-1} \iota_{\partial_{\tau^i}} \right) \tilde{\mathbb{Y}}^{(n-1)} 
\end{equation}
and check that $Q$ in \eqref{TCFE} is conserved. 
As a guiding example, we can explicitly verify it in the simple case of the plane $z=0$ in $\mathbb{R}^3$. We have the following chain of identities
\begin{eqnarray}
\label{TCGA}
\partial_0 Q &=&  \int_{{\cal M}^{(3)} \times \Delta} \partial_0 J_0 \, d^2 x \wedge \tilde{\mathbb{Y}}^{(3)} = 
  \int_{{\cal M}^{(3)} \times \Delta} \star_{\tilde{g}} d (\star_{\tilde{g}} J^{(1)}) \, d^2 x \wedge \tilde{\mathbb{Y}}^{(3)}  =
  \nonumber \\
  &=&   \int_{{\cal M}^{(3)} \times \Delta} d (\star_{\tilde{g}} J^{(1)}) \delta \left( x - \tau_1 \right) \delta \left( y - \tau_2 \right) \delta \left( z \right)  = \nonumber \\
    &=& \int_{{\cal M}^{(3)} \times \Delta} (\star_{\tilde{g}} J^{(1)}) d \left[ \delta \left( x - \tau_1 \right) \delta \left( y - \tau_2 \right) \delta \left( z \right) \right] =
     \nonumber \\
   &=& \int_{{\cal M}^{(3)} \times \Delta} \left[ dy dz d \tau_1 d \tau_2 J_x - dx dz d \tau_1 d \tau_2 J_y + dx dy d \tau_1 d \tau_2 J_z \right] d \left[ \delta \left( x - \tau_1 \right) \delta \left( y - \tau_2 \right) \delta \left( z \right) \right]
     \nonumber \\    
  &=&    
   \int_{{\cal M}^{(3)} \times \Delta} \Big[ J_x \partial_x \delta \left( x - \tau_1 \right) \delta \left( y - \tau_2 \right) \delta \left( z \right) + J_y \delta \left( x - \tau_1 \right) \partial_y \delta \left( y - \tau_2 \right) \delta \left( z \right) + \nonumber \\
   &+& J_z \delta \left( x - \tau_1 \right) \delta \left( y - \tau_2 \right) \partial_z \delta \left( z \right) \Big] dx dy dz d\tau_1 d \tau_2 = \nonumber \\
   &=&    
   \int_{{\cal M}^{(3)} \times \Delta} \Big[ 2 J_x^y \partial_x \delta \left( x - \tau_1 \right) \partial_y \delta \left( y - \tau_2 \right) \delta \left( z \right) + 2 J_y^z \delta \left( x - \tau_1 \right) \partial_y \delta \left( y - \tau_2 \right) \partial_z \delta \left( z \right) + \nonumber \\
   &+& 2 J_z^x \partial_x \delta \left( x - \tau_1 \right) \delta \left( y - \tau_2 \right) \partial_z \delta \left( z \right) \Big] dx dy dz d\tau_1 d \tau_2 = \nonumber \\
   &=&    
   \int_{{\cal M}^{(3)} \times \Delta} \Big[ 2 J_x^y \partial_{\tau_1} \delta \left( x - \tau_1 \right) \partial_{\tau_2} \delta \left( y - \tau_2 \right) \delta \left( z \right) - 2 J_y^z \delta \left( x - \tau_1 \right) \partial_{\tau_2} \delta \left( y - \tau_2 \right) \partial_z \delta \left( z \right) + \nonumber \\
   &-& 2 J_z^x \partial_{\tau_1} \delta \left( x - \tau_1 \right) \delta \left( y - \tau_2 \right) \partial_z \delta \left( z \right) \Big] dx dy dz d\tau_1 d \tau_2 = 0
  \end{eqnarray}
where we have used $\displaystyle \partial_x \delta \left( x - \tau_1 \right) = - \partial_{\tau_1} \delta \left( x - \tau_1 \right) $, $\displaystyle \partial_y \delta \left( y - \tau_2 \right) = - \partial_{\tau_2} \delta \left( y - \tau_2 \right) $ and, after integration by parts, $\partial_{\tau_1} J^i_j = 0 = \partial_{\tau_2} J^i_j$, since the current does not depend on the parameters of the hypersurface.

This construction can be easily generalized to the case of supermanifolds. It is sufficient to include the parametrization of the spinorial coordinates in the immersion equations $\tau_i \to (x^a(\tau_i) , \theta^\a(\tau_i))$. The rest of the procedure remains the same with the obvious modifications due to the replacement of manifolds with supermanifolds.

\section{Linking number and PCO}\label{app:linking} \setcounter{equation}{0}

In this appendix we first recall the basic definition of \emph{linking number} between two curves in three dimensions and prove that it can be expressed in terms of the PCOs describing the immersion of the two curves (so recovering the formula given in \cite{botttu} for the $S^3$ case). This alternative formulation allows for a straightforward generalization to $n$ dimensions where it defines the linking number between two hypersurfaces. It also allows for a generalisation to super-hypersurfaces in supermanifolds, as we are going to discuss. 

In a three-dimensional manifold ${\cal M}^{(3)}$, we consider two closed (oriented) curves $\g_1$ and $\g_2$ defined by the two sets of equations $\phi_1(\vec{x}) = \rho_1(\vec{x}) =0$ and $\phi_2(\vec{x}) = \rho_2(\vec{x}) =0$. The corresponding PCOs localizing on the two curves read explicitly
\begin{equation}\label{appD:1}
	\mathbb{Y}_{\gamma_1}^{(2)} = d \phi_1 \delta ( \phi_1 ) d \rho_1 \delta ( \rho_1) \ , \qquad  \mathbb{Y}_{\gamma_2}^{(2)}  = d \phi_2 \delta ( \phi_2 ) d \rho_2 \delta ( \rho_2 ) 
\end{equation}	
We note that both of them can be rewritten as 
\begin{equation}\label{appD:2}	
\ \mathbb{Y}_{\gamma_i}^{(2)}  =  \left[ d \Theta ( \phi_i ) \right] d \Theta ( \rho_i ) =  d \left[\Theta ( \phi_i )  d \Theta ( \rho_i )\right] \equiv d \Omega_{\gamma_i}^{(1)}  \, , \qquad i=1,2
\end{equation}
where $\Omega_{\g_i}^{(1)} $ are 1-forms with non-compact support. 

Let us consider the Gauss' formula for the linking number of $\g_1, \g_2$
\begin{equation} \label{Gauss}
	l \left( \gamma_1 , \gamma_2 \right) = \oint_{\gamma_1} \oint_{\gamma_2} \frac{(\vec{x}_1 - \vec{x}_2)}{|| \vec{x}_1 - \vec{x}_2 ||^3} \cdot  d \vec{x}_1 \wedge d \vec{x}_2 
\end{equation}
where $\vec{x}_1$ and $\vec{x}_2 $ are the position vectors on the two loops. We state that this formula can be rephrased in terms of PCO's in (\ref{appD:1}, \ref{appD:2}) according to one of the two equivalent expressions
\begin{eqnarray} \label{Bottu}
	l \left( \gamma_1 , \gamma_2 \right) = \int_{\mathcal{M}^{(3)}} \mathbb{Y}_{\gamma_1}^{(2)}  \wedge \Omega_{\gamma_2}^{(1)} \qquad {\rm or} \qquad 
l \left( \gamma_1 , \gamma_2 \right) 	=  \int_{\mathcal{M}^{(3)}}   \Omega_{\gamma_1}^{(1)} \wedge \mathbb{Y}_{\gamma_2}^{(2)} 
\end{eqnarray}
In order to prove the equivalence between expressions \eqref{Gauss} and \eqref{Bottu},  we first note that the integrand in \eqref{Gauss} (that we denote briefly as $G$) is the Green's function of the $d$ operator. Introducing the laplacian $\Delta = \{d , d^\dagger \}$ we can then  formally write $\displaystyle G = \frac{d^\dagger_1}{\Delta} \text{Vol}_1$, where $\text{Vol}_1 = d^3 x_1 \delta^{(3)} \left( x_1 - x_2 \right)$. 
Moreover, if in \eqref{Gauss} we make use of the PCOs to rewrite the integrals over the closed curves as integrals over the entire manifold $\mathcal{M}$, we obtain the following chain of identities
\begin{eqnarray}
l(\g_1, \g_2)   &=& \oint_{\gamma_1} \oint_{\gamma_2} \frac{d^\dagger_1}{\Delta} \text{Vol}_1 = \int_{\mathcal{M}^{(3)}} \int_{\mathcal{M}^{(3)}} \frac{d^\dagger_1}{\Delta} \text{Vol}_1 \; \;\mathbb{Y}_{\gamma_1}^{(2)} \wedge d_2 \Omega_{\gamma_2}^{(1)} \nonumber\\
&=& - \int_{\mathcal{M}^{(3)}} \int_{\mathcal{M}^{(3)}} \frac{d_2 d_1^\dagger}{\Delta} \text{Vol}_1 \; \; \mathbb{Y}_{\gamma_1}^{(2)} \wedge \Omega_{\gamma_2}^{(1)} 
= \int_{\mathcal{M}^{(3)}} \int_{\mathcal{M}^{(3)}} \frac{d_1 d_1^\dagger}{\Delta} \text{Vol}_1 \; \; \mathbb{Y}_{\gamma_1}^{(2)} \wedge \Omega_{\gamma_2}^{(1)} \nonumber\\
&=& \int_{\mathcal{M}^{(3)}} \mathbb{Y}_{\gamma_1}^{(2)} \wedge \Omega_{\gamma_2}^{(1)}
\end{eqnarray}
readily leading to the first expression in \eqref{Bottu}. Writing $\mathbb{Y}_{\gamma_1}^{(2)} = d \Omega_{\gamma_1}^{(1)}$ and integrating by parts, we obtain the second expression in \eqref{Bottu}.

We have considered the particular case of two intertwined lines. However, it is easy to realize that a non-trivial linking number could arise also between a point $P =
(x_{1,P}, x_{2,P},x_{3,P})$  and a surface $\s$ embedded by $\phi(\vec{x}) = 0$. In fact, assigned the corresponding PCOs
\begin{eqnarray}
	\nonumber \mathbb{Y}_P^{(3)} &=& dx^1 \delta ( x_1 - x_{1,P} ) dx^2 \delta ( x_2 - x_{2,P} ) dx^3 \delta ( x_3 - x_{3,P} )  \\
	\mathbb{Y}_\s^{(1)} &=& d \phi \delta ( \phi ) = d \Theta ( \phi ) \equiv d \Omega_\s^{(0)}
\end{eqnarray}
the linking number is a well-defined three-dimensional integral of a 3-form, whose value is 
\begin{equation} 
	l \left( P , \s \right) = \int_{\mathcal{M}^{(3)}} \mathbb{Y}_{P}^{(3)}  \wedge \Omega_\s^{(0)}  
= \int_{\mathcal{M}^{(3)}} d^3x \, \delta^{(3)} ( x - x_P ) \, \Theta ( \phi ) = \begin{cases}
		1 \ , \ \text{if} \; P \in \s \\
		0 \ , \ \text{if} \; P \notin \s
	\end{cases}
\end{equation}

This formulation straightforwardly applies to higher dimensional cases and provides constraints on the possible pairs of submanifolds  which can link non-trivially. 

In a generic $n$-dimensional manifold ${\cal M}^{(n)}$ we consider two hypersurfaces $\s_1, \s_2$ of dimension $(n-p_1)$ and $(n-p_2)$ respectively, with embedding equations $\phi_k(x) = 0$, $k=1, \dots , p_1$ and $\psi_k(x) = 0$, $k=1, \dots , p_2$. The corresponding PCOs localizing on the two submanifolds are then $p_1$- and $p_2$-forms, given by 
\begin{equation}
\mathbb{Y}_{\s_1}^{(p_1)} = \prod_{k=1}^{p_1} d \phi_k \delta ( \phi_k ) \equiv d \Omega_{\s_1}^{(p_1-1)} \; , \qquad \mathbb{Y}_{\s_2}^{(p_2)} = \prod_{k=1}^{p_2} d \psi_k \delta ( \psi_k ) \equiv d \Omega_{\s_2}^{(p_2-1)}
\end{equation}	
where, as for the three-dimensional case, we have written one of the delta functions as the derivative of the Heaviside step function and pulled out the differential. Therefore, the linking number is defined as
\begin{eqnarray}\label{app:general}
\hspace{-1cm} 	l \left( \s_1 , \s_2 \right) = \int_{\mathcal{M}^{(n)}} \mathbb{Y}_{\s_1}^{(p_1)}  \wedge \Omega_{\s_2}^{(p_2-1)} \qquad {\rm or} \qquad 
l \left( \s_1 , \s_2 \right) 	=   \int_{\mathcal{M}^{(n)}}   \Omega_{\s_1}^{(p_1-1)} \wedge \mathbb{Y}_{\s_2}^{(p_2)} 
\end{eqnarray}
This expression is non-vanishing if and only if $p_1 + p_2 = n+1$. Therefore, assigned the dimension of the manifold, this constraint selects which are the dimensions of submanifolds that can actually intertwine. 
For example, in four dimensions we have $(p_1 + p_2)=5$ and the two consistent cases of linkable objects are the case of a point and a three-volume ($p_1=4, p_2 = 1$) and the case of a line and a surface ($p_1=3, p_2 = 2$).
 
\vskip 10pt 
We now generalize definition \eqref{app:general} to the case of a $(n|m)$-dimensional supermanifold. Given a purely bosonic $(n-p_1|0)$-dimensional  hypersurface $\S_1$ and a $(n-p_2|m)$-dimensional super-hypersurface $\Gamma_2$, the corresponding PCOs are\footnote{The definition of super linking number could be extended to the case of generic pseudo-surfaces of dimensions $(p|q)$ with $0 < q < m$. However, since in the body of the paper we only deal with bosonic surfaces ($q=0)$ and supersurfaces ($q=m$), here we stick only to these two cases. }
\begin{eqnarray}\label{appD:super}
\hspace{-0.5cm} \mathbb{Y}_{\S_1}^{(p_1|m)} = \mathbb{Y}^{(p_1|0)}_{\S_1} \prod_{\a=1}^m \theta^\a \delta ( d \theta^\a )  =  
d \Omega_{\S_1}^{(p_1-1|m)}\; , \qquad 
\mathbb{Y}_{\S_2}^{(p_2|0)}  = d \Omega_{\S_2}^{(p_2-1|0)}
\end{eqnarray}
where, as before, the right hand side is obtained by writing one delta function as the derivative of the step function and pulling out the differential. 

Generalizing the previous construction, the \emph{super-linking number} is defined as
\begin{equation}\label{appD:super2}
	L(\S_1, \S_2) = \int_{\mathcal{SM}^{(n|m)}} \mathbb{Y}_{\S_1}^{(p_1|m)} \wedge \Omega_{\S_2}^{(p_2-1|0)} =  \int_{\mathcal{SM}^{(n|m)}} 
	\Omega_{\S_1}^{(p_1-1|m)} \wedge \mathbb{Y}_{\S_2}^{(p_2|0)}
	\end{equation}
We note that these integrals are well-defined only if the bosonic dimensions satisfy $p_1+p_2=n+1$. Instead, the sum of the corresponding odd dimensions already saturates $m$, having chosen from the very beginning  to link a bosonic surface (odd dimension zero) with a super-hypersurface (odd dimension $m$). 

Whenever the sum of the odd dimensions of the two hypersurfaces does not equal $m$, the super-linking number is zero even if the bosonic dimensions sum up to $(n+1)$. For instance two purely bosonic hypersufaces whose bosonic dimensions satisfy the constraint would anyway have super linking number equal to zero. This means that they can be somehow unlinked \virgolette deforming them in the fermionic directions". 

If, instead, the odd dimensions sum up to $m$, the super-linking number \eqref{appD:super2} is well-defined and possibly non-vanishing. We note that, thanks to the particular structure of the PCOs, it actually reduces to the ordinary linking number. In fact, taking for instance the case in \eqref{appD:super}, we find
\begin{equation}
	\int_{\mathcal{SM}^{(n|m)}} \mathbb{Y}^{(p_1|0)}_{\S_1} \prod_{\a=1}^m \theta^\a \delta ( d \theta^\a )  \wedge \Omega_{\S_2}^{(p_2-1|0)}  = \int_{\mathcal{M}^{(n)} \hookrightarrow \mathcal{SM}^{(n|m)}} \mathbb{Y}^{(p_1|0)}_{\S_1} \wedge \Omega_{\S_2}^{(p_2-1|0)}
\end{equation}
The same behavior can be detected in any case. This is somehow not surprising, since the linking number is related to the topological nature of the two hypersurfaces and the fermionic sector never affects topology. 

\newpage

\end{document}